\documentclass[journal]{IEEEtran}

\usepackage{times}
\usepackage{helvet}
\usepackage{courier}
\usepackage{latexsym,bm,amsmath,amssymb,epstopdf}
\usepackage{algorithm}
\usepackage{algorithmic}
\usepackage{graphicx}
\usepackage{subfigure}
\usepackage{color}
\usepackage{colortbl}
\usepackage{array}
\usepackage{caption}
\usepackage{cite}
\usepackage[normalem]{ulem}

\newtheorem{theorem}{\textbf{Theorem}}
\newtheorem{definition}{\textbf{Definition}}
\newtheorem{lemma}{\textbf{Lemma}}
\newenvironment{Proof}{\textit{Proof:}}{\hfill$\square$}

\begin{document}

\title{Privacy-Preserving Collaborative Deep Learning with Unreliable Participants}

\author{Lingchen~Zhao,
		Qian~Wang,~\IEEEmembership{Senior Member,~IEEE,}
		Qin~Zou,~\IEEEmembership{Senior Member,~IEEE,}\\
        Yan~Zhang,
        and Yanjiao~Chen,~\IEEEmembership{Member,~IEEE}

\IEEEcompsocitemizethanks{
\IEEEcompsocthanksitem Qian Wang's research is supported in part by the NSFC under Grants 61822207 and U1636219, the Equipment Pre-Research Joint Fund of Ministry of Education of China (Youth Talent) under Grant 6141A02033327, the Outstanding Youth Foundation of Hubei Province under Grant 2017CFA047, and the Fundamental Research Funds for the Central Universities under Grant 2042019kf0210.
Qin Zou's research is supported in part by the NSFC under Grants 61872277 and 41571437, the  Natural Science Foundation of Hubei Province under Grant 2018CFB482.
Yanjiao Chen's research is supported by the NSFC under Grant 61702380, the Natural Science Foundation of Hubei Province under Grant 2017CFB134, the Hubei Provincial Technological Innovation Special Funding Major Projects under Grant 2017AAA125. (\emph{Corresponding author: Qin Zou})
\IEEEcompsocthanksitem L. Zhao and Q. Wang are with the School of Cyber Science and Engineering, Wuhan University, Wuhan 430072, Hubei, China, and also with The State Key Laboratory of Cryptography, P.O. Box 5159, Beijing 100878, China. Email: \{lczhaocs, qianwang\}@whu.edu.cn.
\IEEEcompsocthanksitem Q. Zou and Y. Chen are with the School of Computer Science, Wuhan University, Wuhan 430072, Hubei, China. Email: \{qzou, chenyanjiao\}@whu.edu.cn.
\IEEEcompsocthanksitem Y. Zhang is with the Huawei Technologies Co., Ltd, Shenzhen 518129, Guangdong, China. Email: zhangyan113@huawei.com. Work done primarily at the School of Computer Science, Wuhan University, China
}}

\markboth{IEEE TRANSACTIONS ON INFORMATION FORENSICS AND SECURITY}
{Shell \MakeLowercase{\textit{et al.}}: Bare Demo of IEEEtran.cls for IEEE Journals}

\maketitle

\begin{abstract}
With powerful parallel computing GPUs and massive user data, neural-network-based deep learning can well exert its strong power in problem modeling and solving, and has archived great success in many applications such as image classification, speech recognition and machine translation etc. While deep learning has been increasingly popular, the problem of privacy leakage becomes more and more urgent. Given the fact that the training data may contain highly sensitive information, e.g., personal medical records, directly sharing them among the users (i.e., participants) or centrally storing them in one single location may pose a considerable threat to user privacy.

In this paper, we present a practical privacy-preserving collaborative deep learning system that allows users  to cooperatively build a collective deep learning model with data of all participants, without direct data sharing and central data storage. In our system, each participant trains a local model with their own data and only shares model parameters with the others. To further avoid potential privacy leakage from sharing model parameters, we use functional mechanism to perturb the objective function of the neural network in the training process to achieve $\epsilon$-differential privacy. In particular, for the first time, we consider the existence of~\textit{unreliable participants}, i.e., the participants with low-quality data, and propose a solution to reduce the impact of these participants while protecting their privacy. We evaluate the performance of our system on two well-known real-world datasets for regression and classification tasks. The results demonstrate that the proposed system is robust against unreliable participants, and achieves high accuracy close to the model trained in a traditional centralized manner while ensuring rigorous privacy protection.

\end{abstract}
\begin{IEEEkeywords}
Collaborative learning, deep learning, privacy
\end{IEEEkeywords}

\section{Introduction}\label{sec:introduction}
\IEEEPARstart{I}{n} the past few years, deep learning has demonstrated largely improved performance over traditional
machine-learning methods in various applications, e.g., image understanding~\cite{he2016deep,zou2019deepcrack}, speech recognition~\cite{graves2013speech,hinton2012deep}, cancer analysis~\cite{fakoor2013using,liang2015integrative}, and the game of Go~\cite{silver2016mastering}. The great success of deep learning is owing to the development of powerful computing processor and the availability of massive data for training the neural networks. In general, the deep learning model will be more accurate if trained with more diverse data, and it motivates companies and institutions to collect as much data as possible from their users. These data are usually generated by sensors on users' personal devices, e.g., GPS, cameras,
smartphones, and heart rate sensors~\cite{ZhouWRKSC19} etc. From the perspective of privacy, however, user-generated data is usually highly sensitive, e.g., location information, personal medical records, and social relationships etc. To gather these sensitive data at a centralized location will raise serious concerns about privacy leakage. A recent regulation of EU~\cite{General} stipulates that companies should carefully collect and use users' personal data, and users have the right to require the company to permanently `forget' their data. The bill also prohibits any \emph{automated individual decision-making} (e.g., personal financial situation, personal health condition, and location prediction) based on the data, and it may greatly affect the machine-learning tasks performed by companies. In addition, in many sectors especially medical industry, sharing personal data is forbidden by laws or regulations. %Therefore, the parallel or distributed methods of machine learning has become a rising concern
%recently, where multiple parties host some of the data, contribute their computing capacities and finally learn a collective machine learning model which benefits from all parties.

To gain the benefit of machine learning while protecting the user privacy, there is a rising interest in designing privacy-assured machine learning algorithms from both academia and industry. Existing solutions for traditional machine learning algorithms mainly exploit intrinsic features of the algorithms, e.g., strict convex objective functions.  Privacy-preserving techniques such as secure multi-party computation or differential privacy have been applied to linear and logistic regression analysis~\cite{du2004privacy, chaudhuri2009privacy}, k-means clustering~\cite{jagannathan2005privacy}, support vector machines~\cite{rubinstein2009learning}, and crowd machine learning~\cite{hamm2015crowd}. In recent years, privacy-preserving deep learning has received much attention from the research community. In~\cite{li2017privynet}, in order to hide the private data, only the intermediate representations obtained by a local neural network are published. However, this scheme did not provide a rigorous privacy guarantee since some sensitive information can be inferred from intermediate features. In~\cite{gilad2016cryptonets}, homomorphic encryption was first applied to convolutional neural networks (CNNs), where the model was trained in a centralized manner, and it required extensive computation resources. Subsequently, many other works tried to make inferences on encrypted data, e.g.,~\cite{liu2017oblivious, mohassel2017secureml, WangDu18, riazi2018chameleon, riazi2019xonn}, etc. However, although the recent schemes have improved the efficiency significantly, they often lead to much higher overheads as compared to computing on the original plaintext data.

To enhance the effectiveness and efficiency of privacy-preserving machine learning approaches, a number of differential privacy based methods have been proposed. In~\cite{abadi2016deep}, a differentially-private stochastic gradient descent algorithm together with a mechanism to accurately track the privacy loss during training were designed, which could train deep neural networks with a modest privacy budget and a manageable model quality. But the scheme still depends on the number of training epochs and some empirical parameters (e.g., the lot size, and the clipping bound). In~\cite{phan2016differential}, differential privacy was applied to a specific deep learning model--deep auto-encoder, and sensitivity analysis and noise insertion were conducted on data reconstruction and cross-entropy error objective functions. In~\cite{shokri2015privacy}, a framework called DSSGD was proposed to ensure differential privacy for distributed deep networks. In recent years, some attacks try to extract alternative information from the machine learning process, e.g., model inversion attacks~\cite{fredrikson2015model}, membership inference attacks~\cite{shokri2017membership}, and model extraction attacks~\cite{tramer2016stealing}. In particular, by utilizing generative adversarial network (GAN), the authors in~\cite{hitaj2017deep} claimed that a distributed deep learning approach cannot protect the training sets of honest participants even if the model is trained in a privacy-preserving manner like~\cite{shokri2015privacy}. However, it was declared in~\cite{GANattack} that the scheme in~\cite{hitaj2017deep} cannot truly break the rigorous differential privacy.

In this paper, we investigate the problem of collaborative deep learning with strong privacy protection while maintaining a high data utility. In our model, users, i.e., participants,
cooperatively learn a collective deep learning model that can benefit from the data of all users. Our work is most related to~\cite{shokri2015privacy, abadi2016deep, phan2016differential, geyer2017differentially}, but is quite
different in several aspects. The proposed schemes in~\cite{abadi2016deep} and~\cite{phan2016differential} were not designed for the collaborative deep learning setting. In~\cite{shokri2015privacy}, participants only shared a subset of parameters with the others to reduce communication costs, and differential privacy is achieved by inserting noises to truncated weights. All the above solutions have their own limitations. The consumed privacy budget during the learning process is relatively high for every single parameter. The total privacy budget is proportional to the number of parameters, which may be in the tens of thousands in deep learning models. The parameter that tunes the fraction of uploaded gradients is used to quantify the privacy, but each pixel in the training data may be revealed by multiple gradients. Compared with~\cite{shokri2015privacy}, the scheme in~\cite{geyer2017differentially} considers providing client-level differential privacy which can hide the existence of participants, and uses the moment accountant technique in~\cite{abadi2016deep} to track the privacy loss. However, both methods did not consider the scenario that the data quality of certain participants may be poor, which may degrade the performance of collaborative learning.

To address the above issues, in our design, each participant, e.g., a mobile user or medical institution, maintains a local neural network model and a local dataset that may be highly sensitive. Instead of sharing local data with the central server, the participant only uploads the updated parameters of the local model generated from the local dataset. The central server derives the global parameters for the collective model using the updates from all participants. Although parameter sharing can prevent direct exposure of the local data, the information of sensitive data may be indirectly disclosed. To solve this problem, we utilize differential privacy~\cite{Dwork06diffpri} to obtain the sanitized parameters to minimize the privacy leakage. Unlike~\cite{shokri2015privacy} where noise is directly injected to the gradients, we apply functional mechanism~\cite{zhang2012functional} to perturb the objective function of the neural network, and obtain the sanitized parameters by minimizing the perturbed objective function.

In collaborative learning, the quality of data contributed by different participants may be very diverse. Different terminal devices or people may have different capacities to generate the training data, and there may exist unpredictable random errors during data collection and storage. Participants with low quality data are referred to as \emph{unreliable participants} (discussed in detail in Section~\ref{sec:problem}). To make the learning process fair and non-discriminative, we consider the `data quality' as one of the privacy concerns of participants, which should not be inferred by other participants during the learning process. We adopt exponential mechanism to protect this privacy while effectively learning an accurate model. Our main contributions are summarized as follows.

\begin{itemize}
\item To the best of our knowledge, we are the first to investigate the problem of privacy-preserving collaborative deep learning by taking into account the existence of unreliable participants.
\item We present a novel scheme called SecProbe, which allows participants to share model parameters, and deals with unreliable participants by utilizing exponential mechanism. SecProbe can protect the data privacy for each participant while effectively learning an accurate model.
\item We derive the approximate polynomial form of the objective function in a neural network with two different loss functions, and use functional mechanism to inject noises to the coefficients to achieve differential privacy without consuming too much privacy budget. We show that it is easy to extend and apply our method to networks with more layers.
\item We evaluate the performance of SecProbe on two well-known real-world datasets for regression and classification tasks. The results demonstrate that SecProbe is robust to unreliable participants, and achieves high accuracy close to the solution obtained in the traditional centralize-trained model, while providing a rigorous privacy guarantee.
\end{itemize}

%The remainder of the paper is organized as follows. Section~ref{secpre} discusses the existing literature on privacy-preserving data publishing techniques and introduces some preliminary knowledge of differential privacy.
%Section~ref{secproblem} gives the problem statement. We
%present SecProbe and analyze its privacy in Section~ref{secSecProbe} and evaluate the performance of the proposed schemes with extensive experiments in Section~ref{secevaluation}. Finally, we conclude the paper in Section~ref{secconclusion}.

%section{Related Work}

\section{Problem Statement and Preliminaries}\label{sec:pre}

In this section, we introduce the problem statement, some preliminary knowledge of deep learning, differential privacy and functional mechanism used in our design.

\subsection{Problem Statement}\label{sec:problem}

\begin{figure}[t!]
  \centering
  \includegraphics[width=0.95\columnwidth]{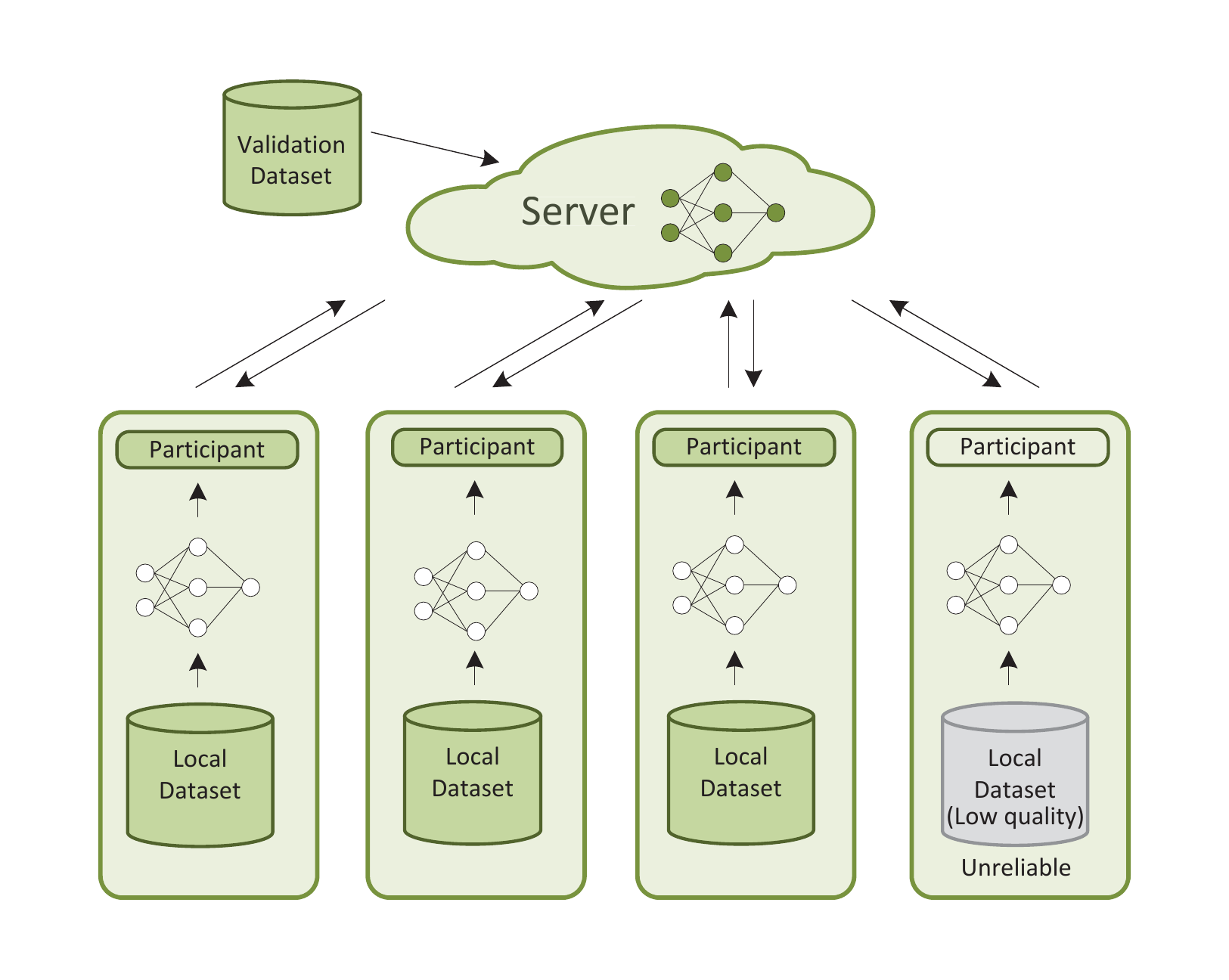}
  %\vspace{-2mm}
  \caption{A collaborative deep learning system with unreliable participants}\label{fig:model}
  %\vspace{-4mm}
\end{figure}

In this paper, we consider the generic setting of privacy-preserving distributed collaborative deep learning. As shown in Figure~\ref{fig:model}, in our model, each participant may have its own sensitive data, and it would like to learn a model benefiting from both its own data and those of the others.

In particular, instead of making the assumption that all the participants are `\textit{reliable}', i.e., the data held by each participant is balanced and has the same or similar quality, we consider a more practical model where there may exist a small group of participants who are `\textit{unreliable}' during some phases of the whole learning process. That is, a portion of data held by unreliable participants is not always as accurate as data held by others, and thus their uploaded parameters may disturb the learning accuracy.

In our daily lives, unreliable participants are common in a collaborative learning system. Consider a typical scenario where several hospitals aim to learn a model together for cancer prediction for patients. There may exist non-negligible gaps in the quality of data among different hospitals since a rich-experienced chief physician with advanced medical devices in a high-rate hospital will be more likely to produce accurate data than an junior physician with low-end of devices in an ordinary hospital. Note that, it does not mean that the data in the ordinary hospital are all and/or always `bad'. Actually, every participant might have some bad data in some phases of their training process when more and more data are being gathered into their local dataset, because there are so many possibilities to go wrong in the data generation and storage procedures. Consequently, the existence of unreliable participants will bring non-ignorable disturbance during the collaborative training process, which may finally result in an inaccurate or even useless model.

Therefore, the problem is how to design a privacy-preserving collaborative deep learning system, which can reduce the impact of unreliable participants on the learned model, and avoid the leakage of participants' privacy from the training data and the trained models.

\emph{Threat Model.}
In our scheme, we assume that the server is \emph{honest-but-curious} and \emph{non-colluding}, i.e., it always follows the protocol, and will not reveal the information of participants or collude with any participant. The participants might be malicious, each participant (or a set of colluding participants) might try to infer other participants' local data and their data quality from the trained model, or upload false parameters/data deliberately into the system.

In addition, to achieve stronger security, we will discuss how to deal with the untrusted server by introducing anonymity communication techniques in Section~\ref{sec:analysis}.

\subsection{Deep Learning}
Broadly speaking, deep learning, based on artificial neural networks, aims to learn and extract high-level abstractions in data and build a network model to describe accurate relations between inputs and outputs. Common deep learning models are usually constructed by multi-layer networks, where non-linear functions are embedded, so that more complicated underlying features and relations can be learned in different layers. Interested readers can refer to thorough surveys or reviews in \cite{schmidhuber2015deep,deng2014tutorial}.

\begin{figure}[t!]
  \centering
  \includegraphics[width=0.85\columnwidth]{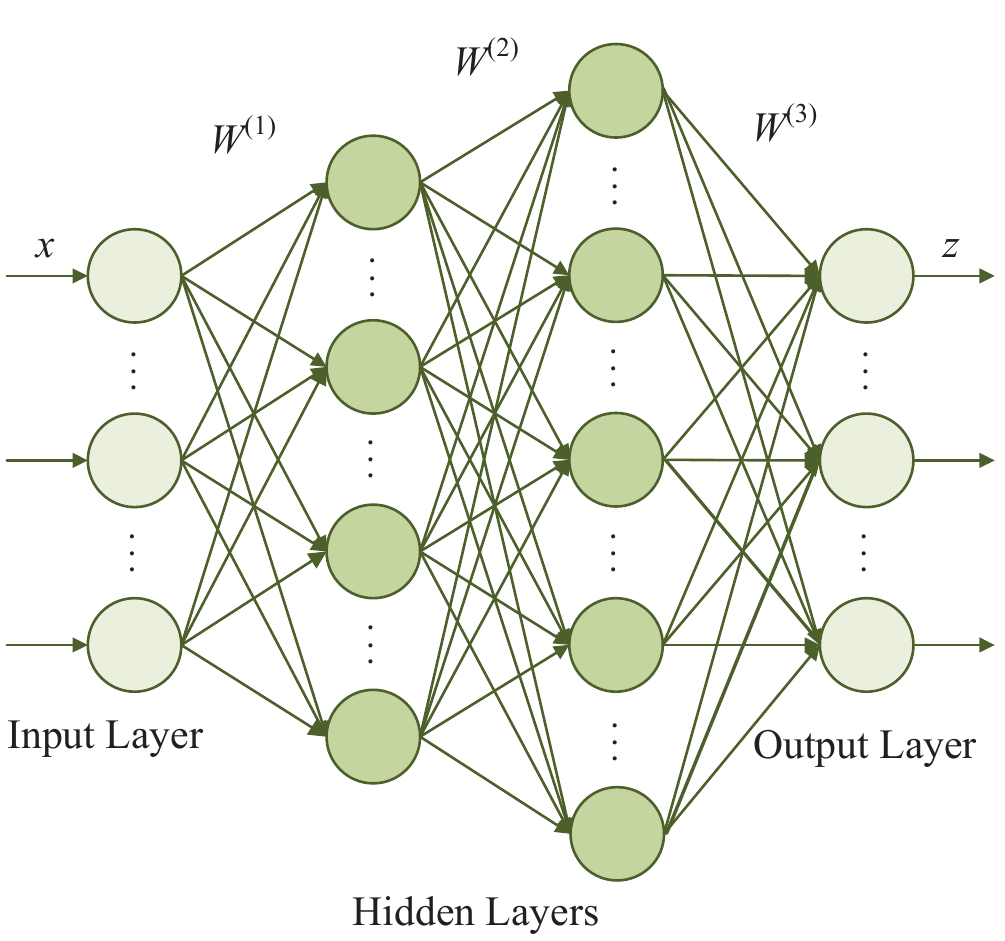}
  %\vspace{-2mm}
  \caption{A neural network with multiple hidden layers}\label{fig:related}
  %\vspace{-4mm}
\end{figure}

There are multiple forms of deep learning models, e.g., multi-layer perceptron (MLP), convolutional neural network (CNN), recurrent neural network (RNN)~\cite{zou2018deep}. Different models fit for different types of problems, and among all of those models, MLP is a very common and representative form of deep learning architecture. Specifically, MLP is a kind of feed-forward neural network, where each neuron receives the outputs of neurons from the previous layer. Figure~\ref{fig:related} shows a typical MLP with multiple hidden layers. Each neuron has an activation function which is usually non-linear. As shown, for a neuron in a hidden layer, say $j$, the output of the neuron is calculated by $h_j = f(W^{(1)}_jx)$, where $W^{(1)}_j$ is the weight vector which determines the contribution of each input signal to the neuron $j$, $x$ is the input of the model, and $f$ is the activation function. The activation function is usually non-linear, capturing the complicated non-linear relation between the output and input. Typical examples are sigmoid function $f(x) = (1 + e^{-x})^{-1}$, \textsf{ReLU} function $f(x) = \max(0, x)$, and hyperbolic tangent $f(x) = \frac{e^{2x} - 1}{e^{2x} + 1}$. In this work, we will focus on an MLP model, where \textsf{ReLU} function is applied.

%\emph{Training neural networks.}
Training a neural network, i.e., learning the parameters (weights) of the network, is a non-convex optimization problem. The typical algorithms used to solve the problem are different types of gradient descent methods~\cite{avriel2003nonlinear}. In this paper, we will consider a supervised learning task, e.g., regression analysis, and assume the output of the network is $z$. Suppose the data we use to train the network is a tuple $(x_i, y_i)$, where $x_i$ denotes the network input and $y_i$ is the label. Consequently, we can use \emph{loss (objective) function} to measure the difference between the network output and the real training label, e.g., $Error_i = (z_i - y_i)^2$. We then can use back propagation \cite{rumelhart1988learning} algorithm to propagate the error back to the neurons, compute the contribution of each neuron to this error, and adjust the weights accordingly to reduce the training error. The adjustment procedure for a weight, say $w_j$, is $w_j = w_j - l\frac{\partial Error_i}{\partial w_j}$, where $l$ is the learning rate.

Among various gradient descent algorithms, \emph{stochastic gradient descent} (SGD)~\cite{zhang2004solving} is considered to be especially fit for optimizing highly non-convex problems for its high efficiency and effectiveness. This algorithm brings stochastic factors into the training process, which helps the model to escape from local optimum. For a large training dataset, SGD first randomly samples a small subset (mini-batch) of the whole dataset, then computes the gradients over the mini-batch and updates the weights, e.g., for weight $w_j$. After one iteration on the mini-batch, the new $w_j$ is computed by $w_j = w_j - l\frac{\partial Error_b}{\partial w_j}$, where $Error_b$ is the loss function computed on the mini-batch $b$. In our work, we will apply SGD to each participant to train its local model.

\subsection{Differential Privacy}
Differential privacy has become a \textit{de facto} standard privacy model for statistics analysis with provable privacy guarantee, and has been widely used in data publishing~\cite{WangZLWQR18, chen2015differentially} and data analysis~\cite{yang2015bayesian, proserpio2014calibrating}. Intuitively, a mechanism satisfies differential privacy if its outputs are approximately the same even if a single record in the dataset is arbitrarily changed, so that an adversary infers no more information from the outputs about the record owner than from the dataset where the record is absent.

\begin{definition}[\textbf{Differential Privacy}~\cite{Dwork06diffpri}]
A privacy mechanism $\mathcal{M}$ gives $\epsilon$-differential privacy, where $\epsilon>0$, if for any datasets $D$ and $D^{'}$ differing on at most one record, and for all sets $S \subseteq Range(\mathcal{M})$,
\begin{equation}\label{eq:differency_privacy}
\centering
\Pr[\mathcal{M}(D)\in S] \leq \exp(\epsilon) \cdot \Pr[\mathcal{M}(D^{'})\in S],
\end{equation}
\end{definition}
where $\epsilon$ is the \emph{privacy budget} representing the privacy level the mechanism provides. Generally speaking, a smaller $\epsilon$ guarantees a stronger privacy level, but also requires a larger perturbation noise.

\begin{definition}[\textbf{Sensitivity}~\cite{DwMc06}]
For any function $f: \mathcal{D} \to \mathcal{R}^d$, the sensitivity of $f$ w.r.t. $\mathcal{D}$ is
\begin{equation}
\Delta(f) = \max_{D, D^{'} \in \mathcal{D}}||f(D)-f(D^{'})||_1
\end{equation}
for all $D$ and $D^{'}$ differing on at most one record.
\end{definition}

Laplace mechanism is the most commonly used mechanism that satisfies $\epsilon$-differential privacy. Its main idea is to add noise drawn from a Laplace distribution into the datasets to be published.

\begin{theorem}[\textbf{Laplace Mechanism}~\cite{DwMc06}]\label{thm:laplace_mechanism}
For any function $f: \mathcal{D} \to \mathcal{R}^d$, the Laplace Mechanism $\mathcal{M}$ for any dataset $D\in\mathcal{D}$,
\begin{equation}
\centering
   \mathcal{M}(D) = f(D) + \langle Lap(\Delta(f) / \epsilon)\rangle ^d
\end{equation}
satisfies $\epsilon$-differential privacy, where the noise $Lap(\Delta(f)/\epsilon)$ is drawn from a Laplace distribution with mean zero and scale $\Delta(f)/\epsilon$.
\end{theorem}

Obviously, the Laplace mechanism only fits for numeric query. Thus, for the query whose outputs are not numeric,  Mcsherry et al. \cite{mcsherry2007mechanism} proposed Exponential mechanism that selects an output $r$ from the output domain $\mathcal{R}$.
%\sout{
%Given an utility function $u(D, r)$ that calculates a score for each output $r$, the exponential function assigns exponentially larger probabilities of being selected to $r$ with a higher score, such that the final output would be approximately optimum with respect to $u$.}

\begin{theorem}[\textbf{Exponential Mechanism}~\cite{mcsherry2007mechanism}]\label{thm:exponential_mechanism}
Let $\Delta u$ be the sensitivity of the utility function $u$: $({\mathcal  {D}}\times {\mathcal{R}})\rightarrow {\mathbb{R}}$, the mechanism $\mathcal{M}$ for any dataset $D\in\mathcal{D}$,
\begin{equation}
\centering
   \mathcal{M}(D, u) = \text{choose $r \in \mathcal{R}$ with probability} \propto \exp(\frac{\epsilon u(D, r)}{2\Delta u})
\end{equation}
gives $\epsilon$-differential privacy.
\end{theorem}

This theorem implies that the Exponential mechanism can make high utility outputs exponentially more likely at a rate that mainly depends on the utility score such that the final output would be approximately optimum with respect to $u$, and meanwhile provide rigorous privacy guarantee.

The composition properties of differential privacy provide privacy guarantee for a sequence of computations.

\begin{theorem}[\textbf{Sequential Composition}~\cite{mcsherry2009privacy}]\label{thm:sequential}
Let $\mathcal{M}_1$, $\mathcal{M}_2$, $\cdots$, $\mathcal{M}_r$ be a set of mechanisms and each $\mathcal{M}_i$ provides $\epsilon_i$-differential privacy. Let $\mathcal{M}$ be another mechanism that executes $\mathcal{M}_1(D)$, $\cdots$, $\mathcal{M}_r(D)$ using independent randomness for each $\mathcal{M}_i$. Then $\mathcal{M}$ satisfies ($\sum_{i}\epsilon_i$)-differential privacy.
\end{theorem}

\begin{theorem}[\textbf{Parallel Composition}~\cite{mcsherry2009privacy}]\label{thm:parallel}
Let $\mathcal{M}_i$ each provide $\epsilon_i$-differential privacy. A sequence of $\mathcal{M}_i$($D_i$)'s over disjoint datasets $D_i$ provide $\max(\epsilon_i)$-differential privacy.
\end{theorem}

These theorems allow us to distribute the privacy budget among $r$ mechanisms to realize $\epsilon$-differential privacy.

%%An important property of differential privacy is the composition property which provides privacy guarantee for a sequence of computations.
%The composition property of differential privacy provides privacy guarantee for a sequence of computations.

%\begin{theorem}[\textbf{Parallel Composition}\cite{}]\label{thm:parallel}
%Let $\mathcal{M}_i$ each provides $\epsilon_i$-differential privacy. A sequence of $\mathcal{M}_i$($D_i$) over disjoint datasets $D_i$ provides $\max(\epsilon_i)$-differential privacy.
%\end{theorem}

\subsection{Functional Mechanism}
Functional mechanism (FM) \cite{zhang2012functional} is a general framework for regression analysis with differential privacy. It can be seen as an extension of the Laplace mechanism which ensures privacy by perturbing the optimization goal of regression analysis instead of injecting noise directly into the regression results.

A typical regression analysis on dataset $D$ returns a model parameter $\dot{w}$ that minimizes the optimization (objective) function $f_D(w) = \sum_{x_i \in D}f(x_i, w)$. However, directly releasing $\dot{w}$ would raise privacy concern, since the parameters reveal information about dataset $D$ and function $f_D(w)$. In order to achieve differential privacy, we use FM to firstly perturb the objective function $f_D(w)$ (by exploiting the polynomial representation of $f_D(w)$), and then release the parameter $\bar{w}$ that minimizes the perturbed objective function $\bar{f}_D(w)$.
%FM perturbs the objective function by exploiting the polynomial representation of $f_D(w)$.

We assume $w$ is a vector containing $d$ values $w_1, \ldots, w_d$. Let $\phi(w)$ denote the product of $w_1,\ldots, w_d$, i.e., $\phi(w) = w_1^{c_1} \cdot w_2^{c_2}\cdots w_d^{c_d}$, where $c_1, \ldots, c_d \in N$. Let $\Phi_j (j \in N)$ denote the set of all products of $w_1, \ldots, w_d$ with degree $j$, i.e., $\Phi_j = \{w_1^{c_1} \cdot w_2^{c_2}\cdots w_d^{c_d} | \sum_{l=1}^{d}c_l=j\}$. By the Stone-Weierstrass Theorem \cite{rudin1964principles}, any continuous and differentiable function $f(w)$ can always be written as a polynomial of $w_1, \ldots, w_d$, i.e., $f(x_i, w) = \sum_{j=0}^{J}\sum_{\phi \in \Phi_j}\lambda_{\phi_{x_i}}\Phi(w)$, where $\lambda_{\phi_{x_i}} \in R$ denotes the coefficient of $\phi(w)$ in the polynomial, and $J \in [0, \infty]$. Similarly, we can derive the polynomial function of $f_D(w)$ as
\begin{equation}\label{equa:poly dataset}
  f_D(w) = \sum_{j=0}^{J}\sum_{\phi \in \Phi_j}\sum_{x_i \in D} \lambda_{\phi_{x_i}}\Phi(w).
\end{equation}

\begin{lemma}\label{lemma:fm}
(\cite{zhang2012functional}) Let $D$ and $D'$ be any two neighboring databases. Let $f_D(w)$ and $f_D'(w)$ be the objective functions of regression analysis on $D$ and $D'$,  respectively. Then, we have the following inequality

%{\footnotesize
\begin{equation*}\label{equation:sensitivity fm}
\begin{split}
  \Delta &= \sum_{j=1}^{J}\sum_{\phi \in \Phi_j} \bigg \| \sum_{x_i \in D}\lambda_{\phi_{x_i}} - \sum_{x_i \in D'}\lambda_{\phi_{x_i^{'}}} \bigg \|_1  \\
  & \leq 2\max\limits_{x} \sum_{j=1}^{J}\sum_{\phi \in \Phi_j} \| \lambda_{\phi_x} \|_1.
\end{split}
\end{equation*}
%}

\end{lemma}

To achieve $\epsilon$-differential privacy, FM perturbs $f_D(w)$ by injecting Laplace noise into its polynomial coefficients. According to Lemma~\ref{lemma:fm}, $f_D(w)$ is perturbed by injecting Laplace noise with scale of $Lap(\Delta / \epsilon)$ into the polynomial coefficients $\lambda(\phi)$, where $\Delta = 2\max\limits_{x} \sum_{j=1}^{J}\sum_{\phi \in \Phi_j} \| \lambda_{\phi_x} \|_1$. Then we can derive the model parameter $\bar{w}$ which minimizes the perturbed function $\bar{f}_D(w)$. In this work, we propose to utilize FM in our design to protect the privacy of participants' local data.

\section{SecProbe: Privacy-Preserving Collaborative Deep Learning System}\label{sec:SecProbe}
\subsection{System Architecture}

\begin{algorithm}[t!]
\caption{A high-level description of SecProbe}\label{alg:high-level}
%{\footnotesize
\begin{algorithmic}[1]
\STATE Build the models and initialize all parameters
\FOR {each communication round}
\FOR {each participant $i$}
\FOR {iteration $j$ = $1$ to $I$}
\STATE Run SGD independently on the local dataset using the perturbed loss function
\ENDFOR
\STATE Upload $\bar{W_i}$ to the server
\ENDFOR
\STATE The server chooses to accept $\bar{W_i}$ according to the computed utility score.
\STATE The server conducts \emph{model average} to obtain $W_{new}$ and send it to each participant.
\ENDFOR
\end{algorithmic}
%}
\end{algorithm}

In Figure~\ref{fig:model}, we assume there are $N$ participants, and each of them has a sensitive dataset for local training. The participants aim to learn a common model, i.e., the architectures of the local models are identical, and the learning objectives are the same. There are many deficiencies and difficulties of collecting all the data from participants in advance and training on the entire dataset. Such complicated process of data collection usually incurs high communication overhead, and participants may not be willing to directly upload their data to a third party from the perspective of privacy or business consideration. Therefore, the participants only exchange the parameters (weights) with others, and a server (e.g., a cloud service provider) undertakes the job of communicating with participants, exchanging and storing parameters.
In our model, we assume there exists a global model and an auxiliary validation dataset on the server. This dataset can be very small, and it is easy to be obtained in practice. For example, the data can be collected from participants who have already been expired with no privacy concern or publicly well-tested hand-classified datasets  (such as MNIST~\cite{lecun1998gradient}).

Algorithm~\ref{alg:high-level} gives the high-level steps of SecProbe. The server and participants build their own models and initialize all the parameters before the learning starts. For each communication round, the participants locally train their own models using SGD in a differentially private way. After \emph{I} times of iteration, the participants upload the perturbed parameters to the server. The server then uses the auxiliary validation data to compute a utility score for each participant, and then chooses to accept the parameters with certain probability. Next, the averaged model parameters are computed and distributed to each participant for the next round of local training. Note that the participant can terminate its training procedure and drop out from the system at any time if it believes its model is accurate enough, and meanwhile a new participant can also join into the system at anytime. We next describe the detailed procedures of SecProbe on the server side and the participant side, respectively.

\subsection{SecProbe: The Server Part}
Algorithm~\ref{alg:server} gives the pseudocode of SecProbe on the server side. The server first initializes the parameters and waits for the local training results from each participant. When the number of participants who upload their weights to the server reaches a pre-fixed threshold \emph{M}, the server stops receiving the uploaded data and sends a stop signal to notify other participants that there is no need to upload weights (in step 2). The parameter \emph{M} is used to control the number of participants that the server plans to utilize per round and meanwhile saves a lot of communication costs. Alternatively, this procedure can also be achieved by randomly assigning a set of $M$ participants at the beginning of each round. These two approaches have their own advantages and both can be used in our design. The former can intrinsically deal with the occurrence of failed uploads, while the latter can save a lot of computation costs on the participant side. Without loss of generality, we adopt the first approach in the description of Algorithm~\ref{alg:server}.

\begin{algorithm}[t!]
\caption{SecProbe on the server side}
%{\footnotesize
\label{alg:server}
\begin{algorithmic}[1]
\STATE Initialize parameters $W_0$ and send them to each participant
\STATE Wait for participants to upload their weights until there are already $M$ participants' weights $\bar{W_1}, \bar{W_2}, \ldots, \bar{W_M}$
\STATE Calculate the accuracy score $u(G, D,m)$ for each uploading participant by running the model with weights from each of them over the auxiliary validation dataset
\STATE Sample $K$ participants from $M$ without replacement such that $\textbf{Pr}[\text{Selecting participant $m$}] \propto \exp(\frac{\epsilon}{2K\Delta u}u(G, D, m))$
\STATE Average the $K$ weights and obtain $W_{new} = \frac{1}{K}\sum_{i=1}^{K}\bar{W_i}$
\STATE Send the new averaged weights to all participants
\STATE Repeat steps 2-6 until there is no participant in the system
\end{algorithmic}
%}
\end{algorithm}

As discussed above, the existence of unreliable participants indicates that the parameters uploaded by them may be disruptive, and it may reduce the accuracy of the global model. To reduce their effect on the model accuracy, we measure the data quality of these unreliable participants by calculating a utility score for each participant. Specifically, the server runs the model on the auxiliary validation dataset $D$ with the weights of each of \emph{M} participants respectively and obtains a utility score for participant $m$. Let $G=[\bar{W_1},\ldots,\bar{W_m}]$ denote the set of uploaded weights, where each item can be used to infer the data quality of each participant. For a regression task, suppose the dataset has $d$ samples, we define a scoring function $u(G, D, m)$ as
\begin{equation}\label{equ:accuracy score}
  u(G, D, m) = \frac{1}{d}\sum_{i}^{d}(1 - |\frac{\max(z_i,3y_i)-y_i}{y_i}|),
\end{equation}
where $z_i$ is the output of the model with parameter $G(m)$, and $y_i$ is the real value from the auxiliary validation data. Without loss of generality, we assume that $y_i$ is in range $[0, 1]$. The scoring function calculates an accuracy score for each participant $m$.

If the server chooses the participants only according to the scoring function without any uncertainty, the participants could easily infer which participants hold the low-quality data by comparing their own parameters and the new parameters sent from the server. In SecProbe, we utilize the exponential mechanism to inject uncertainty into the sampling procedure against this kind of inference. The server samples $K$ participants without replacement such that
\begin{equation}\label{equ:prob}
  \textbf{Pr}[\text{Selecting participant $m$}] \propto \exp(\frac{\epsilon}{2K\Delta u}u(G, D, m)).
\end{equation}

However, the sensitivity of calculating $1 - |\frac{z_i-y_i}{y_i}|$ is unscalable since the proportion of $z_i$ to $y_i$ is infinite in theory. If $z_i\leq3y_i$, the value will be within $[-1, 1]$. We observe that this condition always holds in practice, as the predicted parameter $z_i$ does not deviate from the real value $y_i$ for more than three times\footnote{We find that the average of $|\frac{z_i-y_i}{y_i}|$ is always less than 1 based on our experimental results in Section~\ref{sec:evaluation}}. To this end, we place a restriction on the output $z_i$, i.e., replacing $z_i$ in Eq.~(\ref{equ:accuracy score}) with $\max(3y_i, z_i)$. If $z_i\le 3y_i$, it will be clipped to avoid the unbounded sensitivity. By adding this restriction, the sensitivity can be bounded to 1. Experimental results show that this design will not affect the accuracy of the learned model.

Moreover, for a classification task, the scoring function $u$ can be defined as the correct prediction rate directly.

\begin{lemma}
 The sensitivity of the prediction accuracy for classification is $\Delta u = \frac{1}{2}$.
\end{lemma}
\begin{Proof}
Let $m$ and $n$ denote the number of correct predictions and the number of samples respectively. The sensitivity is $\Delta = \frac{m+1}{n+1} - \frac{m}{n} = \frac{n-m}{n(n+1)}$. Since $n\geq 1$ and $n\geq m$, the maximum of $\Delta u$ is $\frac{1}{2}$ when $n=1$ and $m=0$.
\end{Proof}

\emph{Remarks. }In this step, the goal is to make the participants in the collaborative learning process non-discriminative. To achieve this, we use the exponential mechanism to prevent adversaries from knowing the performances of uploaded models, i.e., the utility scores. The privacy of the participants' training data will be considered in the next step. As it is unfair to measure the performance of the participants' own data, a reliable dataset is required for evaluation, e.g., a validation dataset on the server. In our design, since the performance of the model is reflected by the prediction accuracy, we protect the prediction values on the validation dataset.
%the preserved data consists of the prediction results on the validation dataset in essence.}
Specifically, the server constructs a centralized virtual dataset which has the same number of records with the original validation dataset, and each record has $M$ attributes which represent the prediction results (the clipped $z_i$ for regression, or the correctness for classification) on the $M$ participants' models. Then, the neighboring dataset consists of two virtual datasets which are different in only one row. Suppose a query asks: ``which model has the best quality?''. This query can be regarded as ``which model gets the highest utility score on the auxiliary validation dataset?''. Obviously, the effect of removing or adding one record in the virtual dataset on the utility score is the sensitivity as above. For the regression task, the summation term in Eq.~(\ref{equ:accuracy score}) is in range [-1, 1]. So, changing one record in the virtual dataset could place an impact on the utility score no larger than 1. For the classification task, adding or removing one record will affect the number of correct predictions. As the neighboring dataset is non-empty, the maximum influence on the utility score is $\frac{1}{2}$.

\begin{theorem}\label{thm:server}
  The sampling procedure in Algorithm~\ref{alg:server} (line 4) satisfies $\epsilon$-differential privacy.
\end{theorem}

\begin{Proof}
\textit{Proof sketch.} Because sampling one participant consumes $\frac{\epsilon}{K}$ budget and satisfies $\frac{\epsilon}{K}$-differential privacy according to Theorem~\ref{thm:exponential_mechanism}, the sampling procedure which samples $K$ participants will satisfy $\frac{K\epsilon}{K}$-differential privacy.
\end{Proof}

\begin{algorithm}[t!]
\caption{SecProbe on the participant side}
\label{alg:participant}
%{\footnotesize
\begin{algorithmic}[1]
\STATE Download the same initialized weights $W_0$ from the server
\STATE Set the mini-batch size $|S|$ and the number of iterations $I$, and perform local SGD in each communication round
\STATE Decompose the loss function $\ell(S, W)$ and derive an approximated polynomial form $\hat{\ell}(S, W)$
\STATE Obtain the perturbed loss function $\bar{\ell}(S, W)$ by functional mechanism
\FOR {iteration $j$ = $1$ to $I$}
\STATE Run SGD with batch size $|S|$ on the local dataset using the perturbed loss function
\ENDFOR
\STATE Upload $\bar{W}$ to the server
\STATE Receive the new averaged weight $W_{new}$ from the server
\STATE Repeat steps 5-9 until an acceptable small test error is obtained
\STATE Drop out of the system
\end{algorithmic}
\end{algorithm}

Note that Theorem~\ref{thm:server} only ensures that this sampling procedure satisfies $\epsilon$-differential privacy at the current training iteration. Due to the composition properties of differential privacy, the privacy level provided by it may degrade during training. We will discuss in detail which privacy level our mechanism will provide during the whole training process in the \emph{privacy analysis} of this section.

\textit{Remarks.} The above procedure samples a set of participants at an exponential rate based on the scoring function while preventing the sampling procedure from leaking privacy. Therefore, the real quality of uploaded weights from a participant cannot be inferred by others since the new weights are computed on a set of privately-chosen participants, and the system can sample the approximately optimal weights and eliminate the disturbance of unreliable participants as much as possible. It is easy to see that the time complexity of the sampling step is $\mathcal{O}(K M)$. We can further significantly reduce the running time by implementing the sampling step on a static balanced binary tree as suggested in~\cite{bhaskar2010discovering}. The improved sampling step can run in time $\mathcal{O}(M + K\ln(M))$.

After choosing the final accepted weights, the server conducts a \emph{model average} operation that sets the new global weights to be the average of all the accepted weights. The server finally sends the new weights to every participant, and waits for the next round of parameter uploading. We now briefly explain the reason why \emph{model average} operation works. The average operation to some extent consistently inherits the procedure of SGD by randomly choosing a mini-batch of the training data to get the sum of errors on the mini-batch and then computing the gradients on the error. The average operation acts as choosing a mini-batch of the data from \emph{all} the accepted participants and computing the gradients on the overall error. Note that, our experiments show that this operation works well only if the parameters of each participant are randomly initialized by the same \textit{seed}, which is easy to be implemented, e.g., the participants can download the same initialized parameters from the server to replace their own initializations at the very beginning.

\begin{figure}[t!]
  \centering
  \includegraphics[width=0.70\columnwidth]{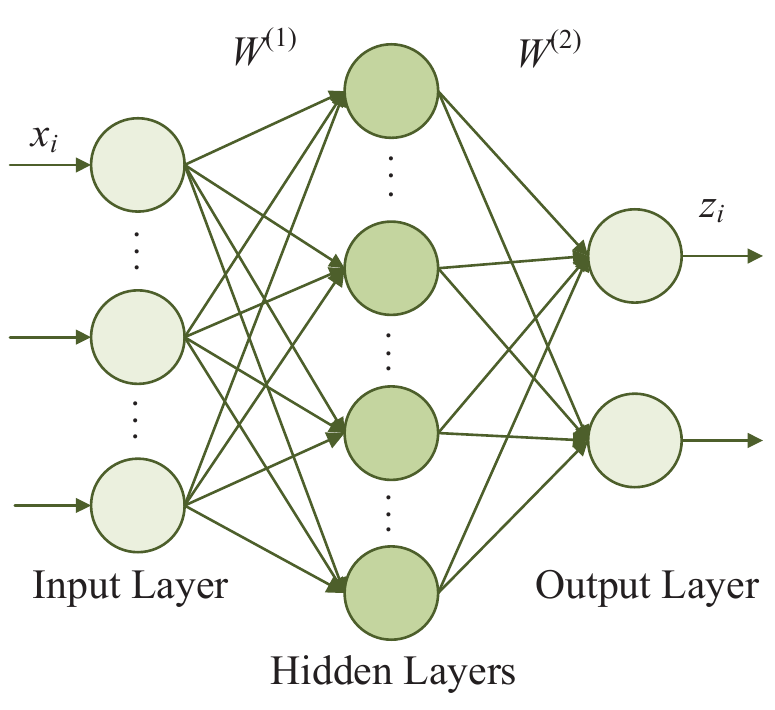}
  %\vspace{-2mm}
  \caption{A neural network with one hidden layer}\label{fig:participant}
  %\vspace{-4mm}
\end{figure}

\subsection{SecProbe: The Participant Part}

Algorithm~\ref{alg:participant} presents the pseudocode of SecProbe on the participant side. Each participant has its own local training dataset and conducts the standard SGD algorithm to train its local model. Let $W_i$ denote the network weights of participant $i$. To protect the privacy of the participant's sensitive data being disclosed by $W_i$, the participant applies differential privacy onto the training algorithm to get sanitized weights $\bar{{W_i}}$'s, and uploads them to the server.

To achieve differential privacy, Laplace mechanism was utilized in~\cite{shokri2015privacy} to directly inject noise to the weights. However, their scheme has to consume too much privacy budget for each weight per epoch in order to achieve acceptable results. Instead of directly injecting noise to the weights $W$, in our design we propose to utilize functional mechanism \cite{zhang2012functional} to perturb the objective function of the network, train the model on the perturbed objective function and finally compute the sanitized weights $\bar{W}$. Since the structures of the neural networks may be varied and often depend on specific application scenarios, it is impossible to design a one-size-fits-all differentially-private solution for all deep learning models. In this paper, we focus on the most common neural network MLP. Specifically, we first consider a three-layer fully-connected neural network, design algorithms to train the model in a differentially-private manner, and then show that more hidden layers can be stacked easily by using our proposed scheme.

The regression problem usually uses \emph{mean square error} (MSE) as the loss function. Suppose the training set $D$ has a set of $n$ tuples $\kappa_1, \kappa_2,  \ldots, \kappa_n$. For each tuple $\kappa_i = (X_i, y_i)$,  $X_i$ contains $d$ attributes $(x_{i1}, x_{i2}, \ldots, x_{id})$ and $y_i$ is the label of $\kappa_i$. Without loss of generality, we assume each attribute in $X_i$ and $y_i$ is in the range $[0,1]$, which is easy to be satisfied by data normalization. The MLP takes $X_i$ as input and outputs a prediction $z_i$ of $y_i$ as accurate as possible. Then, the objective function can be given by
%{\footnotesize
\begin{equation}\label{equ:mlp}
  \ell(D, W) = \sum_{i=1}^{n}(z_i - y_i)^2.
\end{equation}
%}
Recall the calculations of MLP, we have $z_i = \sigma_1(HW^{(2)})~\mathrm{and}~H = \sigma_2(X_i^TW^{(1)})$,
%\begin{equation*}\label{equ:mlp-output}
%\begin{split}
%  \Theta_i = \sigma_1(HW^{(2)})~\mathrm{and}~H = \sigma_2(X_i^TW^{(1)}),
%\end{split}
%\end{equation*}
where $W^{(1)}$ and $W^{(2)}$ are the weight matrixes of the network (as shown in Figure~\ref{fig:participant}), $\sigma_1$ is the sigmoid function, and $\sigma_2$ is the \textsf{ReLU} function. Note that we bound the ReLU function by $[0, 1]$ to avoid introducing an unbounded global sensitivity.

Consequently, for a mini-batch $S$ sampled from the training set $D$, the objective function can be written into the following form
\begin{equation}\label{equ:mlp-loss}
\begin{split}
\ell(S, W) &= \sum_{i=1}^{|S|}(z_i - y_i)^2 \\
 & = \sum_{i=1}^{|S|}[y_i^2 -2y_i[1 +  e^{(-(\textsf{ReLU}(X_i^TW^{(1)})W_j^{(2)}))}]^{-1} \\
 & + [1 + e^{(-(\textsf{ReLU}(X_i^TW^{(1)})W_j^{(2)}))}]^{-2} ].
\end{split}
\end{equation}

%%{\footnotesize
%\begin{equation}\label{equ:mlp-loss}
%\begin{aligned}
%\ell(S, W) =  \sum_{i=1}^{|S|}\sum_{j=1}^{b}(\theta_{ij} - y_{ij})^2= \sum_{i=1}^{|S|}\sum_{j=1}^{b} [y_{ij}^2 -2y_{ij}[1 +~~~~~~~~~~~~~~~~~~~~~~~~~~\\
% e^{(-(\mathrm{ReLU}(X_i^TW^{(1)})W_j^{(2)}))}]^{-1}]
% + [1 + e^{(-(\mathrm{ReLU}(X_i^TW^{(1)})W_j^{(2)}))}]^{-2}.~~~~~~~~~~
%\end{aligned}
%\end{equation}
%%}

Recall that FM requires the objective function to be the polynomial representation of weights $w$, thus we need to approximate Eq.~(\ref{equ:mlp-loss}) and rewrite it into a polynomial form. Since the first term of Eq.~(\ref{equ:mlp-loss}) is already in the polynomial form, we only consider the other two terms. To utilize \emph{Taylor Expansion} to help approximate the functions as suggested in~\cite{zhang2012functional}, $\forall j \in [1, d]$, we define four functions $f_{1j}, f_{2j}, g_{1j}$ and $g_{2j}$ as follows.
\begin{equation}\label{equ:mlp-taylor}
\begin{split}
  f_{1} &= -2y_{i}[1+\exp(-z)]^{-1}; f_{2} = [1+\exp(-z)]^{-2}; \\
  g_{1} &= \textsf{ReLU}(X_i^TW^{(1)})W^{(2)}; g_{2} = \textsf{ReLU}(X_i^TW^{(1)})W^{(2)}.
\end{split}
\end{equation}
Then, we can rewrite Eq.~(\ref{equ:mlp-loss}) into the following form
\begin{equation}\label{equ:mlp-t-single}
\ell(S, W) = \sum_{i=1}^{|S|}[y_{i}^2 + f_{1}(g_{1}(\kappa_i, W)) + f_{2}(g_{2}(\kappa_i, W))].
\end{equation}

\begin{figure*}[h!t]
\begin{eqnarray}\label{equ:mlp-taylor-finite}
\hat{\ell}(S, W)
&=& \sum_{i=1}^{|S|}\!\!
  \Big[y_{i}^2 + \!\! \sum_{l=1}^{2}\sum_{k=0}^{2}\!\!\big[\frac{f_{l}^{(k)}(\gamma_{l})}{k!}\big(g_{l}(\kappa_i, W) \!-\! \gamma_{l}\big)^{k}\big]\!\Big]\! \nonumber\\
 &=& \sum_{i=1}^{|S|}\!\! \Big[\!\big(y_{i}^2\! +\!\! \sum_{l=1}^{2}f_{l}^{(0)}\!(0)\big)\! + \!\sum_{l=1}^{2}f_{l}^{(1)}(0)\!\big(\!\textsf{ReLU}(X_i^TW^{(1)}\!)W^{(2)}\!\big)
  + \sum_{l=1}^{2}\frac{f_{l}^{(2)}(0)}{2}\big(\textsf{ReLU}(X_i^TW^{(1)})W^{(2)}\big)^{2} \Big] \nonumber\\
 &=& \sum_{i=1}^{|S|}\!\! \Big[y_{i}^2 - y_{i} + \frac{1}{4} + \frac{1-2y_{i}}{4}\big(\textsf{ReLU}(X_i^TW^{(1)})W^{(2)}\big) + \frac{1}{16}\big(\textsf{ReLU}(X_i^TW^{(1)})W^{(2)}\big)^2\Big]
\end{eqnarray}
\end{figure*}

Given the above decomposition of the original loss function, we can then apply Taylor expansion in Eq.~(\ref{equ:mlp-t-single}) and obtain Eqs.~(\ref{equ:mlp-taylor-finite}) and (\ref{equ:mlp-taylor-infinite}),
\begin{equation}\label{equ:mlp-taylor-infinite}
  \tilde{\ell}(S, W) = \sum_{i=1}^{|S|}\Big[y_{i}^2 + \sum_{l=1}^{2}\sum_{k=0}^{\infty}\frac{f_{l}^{(k)}(\gamma_{l})}{k!}\big(g_{l}(\kappa_i, W) - \gamma_{l}\big)^{k}\Big],
\end{equation}
where $\gamma_{l}$ is a real number and without loss of generality we set it to be zero for ease of analysis. As can be seen in Eq.~(\ref{equ:mlp-taylor-infinite}), the number of polynomial terms is infinite, which may result in an unacceptable large sensitivity. Thus, we propose to truncate Eq.~(\ref{equ:mlp-taylor-infinite}) by cutting off all polynomial terms with order larger than 2, i.e., we set $k \in [0,2]$. Then we can obtain the final polynomial objective function used for training as Eq.~(\ref{equ:mlp-taylor-finite}). The influence of truncating the polynomial and clipping the ReLU function is shown in Table~\ref{tab:truncation}. We can see that, both operations do not have obvious impacts on the training accuracy.

Now we are ready to give the following lemma.

\begin{lemma}\label{lemma:mlp-sensitivity}
Let $S$ and $S'$ be any two neighboring databases. Let $\hat{\ell}(S, W)$ and $\hat{\ell}(S', W)$ be the objective functions of MLP on $S$ and $S'$ respectively. Then, the global sensitivity of the objective function $\hat{\ell}$ over $S$ and $S'$ is
\begin{equation*}\label{equation:sensitivity-mlp}
\begin{split}
   \Delta & \leq \frac{1}{2}b + \frac{1}{8}b^2,
\end{split}
\end{equation*}
where $b$ is the number of hidden units in the hidden layer.
\end{lemma}

\begin{Proof}
Without loss of generality, we assume that $S$ and $S'$ differ in the last tuple, $\kappa_{|S|} (\kappa_{|S|}^{'})$. According to Lemma~\ref{lemma:fm}, we have
\begin{equation*}\label{equ:fm-proof}
\begin{split}
  \Delta & \leq 2\max_{\kappa}\big(\frac{1}{4}\sum_{p=1}^{b}h_p + \frac{1}{16}\sum_{p=1,q=1}^{b}h_ph_q \big) \\
  &\leq 2(\frac{1}{4}b + \frac{1}{16}b^2) =  \frac{1}{2}b + \frac{1}{8}b^2,
\end{split}
\end{equation*}
where $h$ is the value of hidden neurons at the hidden layer.
\end{Proof}

As can be seen, the sensitivity of the objective function $\hat{\ell}(S, W)$ only depends on the model structure, which is independent with the cardinality of the dataset $S$.
Finally, we inject Laplace noise with scale $\frac{\Delta}{\epsilon}$ to the coefficients of $\hat{\ell}(S, W)$ and obtain the perturbed objective function $\bar{\ell}(S, W)$, which satisfies $\epsilon$-differential privacy.

The classification problem usually adopts \emph{cross-entropy error} as the loss function. We construct a CNN the same as~\cite{shokri2015privacy} which has two convolutional layers, two max pooling layers and one hidden layer with 128 neurons and sigmoid activation function as an example to solve the classification problem. Similar to the regression problem, the objective function is
\begin{equation}\label{equ:cross_entropy}
  \ell(\kappa_i, W) = \sum_{i=1}^n-y_{i}\log z_{i}.
\end{equation}
The Eq.~(\ref{equ:cross_entropy}) can also be decomposed to three functions as follows
\begin{equation*}\label{equ:mlp-cls-taylor}
\begin{split}
  f &= -y_{i}\log[1+\exp(-z)]^{-1}~\mathrm{and} \\
  g_1 &= g_{2} = Conv(\kappa_i)W.
\end{split}
\end{equation*}
Note that, $Conv(\kappa_i)$ represents the output of the previous pooling layer, and $W$ is the weight matrix of the fully-connected layer. Then, the loss function can be rewritten as Eq.~(\ref{equ:mlp-cls-single}),
\begin{equation}\label{equ:mlp-cls-single}
  {\ell}(S, W) = -\sum_{i=1}^{|S|}\sum_{j=1}^Mf((\kappa_i, W)).
\end{equation}
The expansion form of Eq.~(\ref{equ:mlp-cls-single}) is
\begin{equation}\label{equ:mlp-cls-taylor-infinite}
  \tilde{\ell}(S, W) = -\sum_{i=1}^{|S|}
  \sum_{j=1}^M\sum_{k=0}^{\infty}\big[\frac{f_{l}^{(k)}(\gamma_{l})}{k!}\big(g_{l}(\kappa_i, W) - \gamma_{l}\big)^{k}\big].
\end{equation}
The final polynomial objective function used for training is given in Eq.~(\ref{equ:mlp-cls-taylor-finite}).

\begin{figure*}[h!t]
\begin{equation}\label{equ:mlp-cls-taylor-finite}
\hat{\ell}(S, W)
= -\sum_{i=1}^{|S|}
  \sum_{j=1}^{M}\sum_{k=0}^{2}\Big[\frac{f_{l}^{(k)}(\gamma_{l})}{k!}\big(g_{l}(\kappa_i, W) - \gamma_{l}\big)^{k}\Big]
 = \sum_{i=1}^{|S|}\Big[\log 2 + (\frac{1}{2}-y_{i})Conv(\kappa_i)W^{} + \frac{1}{8}[Conv(\kappa_i)W^{}]^2\Big]
\end{equation}
\end{figure*}

\begin{lemma}\label{lemma:mlp-cls-sensitivity}
Let $S$ and $S'$ be any two neighboring databases. Let $\hat{\ell}(S, W)$ and $\hat{\ell}(S', W)$ be the objective functions of MLP on $S$ and $S'$ respectively. The global sensitivity of the objective function $\hat{\ell}$ over $S$ and $S'$ is
\begin{equation*}\label{equation:sensitivity-mlp}
\begin{split}
   \Delta & M(\leq b + \frac{1}{4}b^2).
\end{split}
\end{equation*}
\end{lemma}

\begin{Proof}
Without loss of generality, we assume that $S$ and $S'$ differ in the last tuple, $\kappa_{|S|} (\kappa_{|S|}^{'}).$ According to Lemma~\ref{lemma:fm}, we have
\begin{equation*}\label{equ:fm-cls-proof}
\begin{split}
   \Delta & \leq 2M\max_{\kappa}\big( \frac{1}{2}\sum_{p=1}^{b}h_p + \frac{1}{8}\sum_{p=1,q=1}^{b}h_ph_q \big) \\
  &\leq 2M(\frac{1}{2}b + \frac{1}{8}b^2) =  M(b + \frac{1}{4}b^2).\\
\end{split}
\end{equation*}
\end{Proof}

We next revisit Algorithm~\ref{alg:participant}. After obtaining the perturbed loss function $\bar{\ell}(S, W)$, the participant performs standard SGD algorithm with the batch size $|S|$ and iterates for $I$ times. Then it gets the sanitized weights $\bar{W}$'s of the current round and uploads them to the server. The parameter $I$ manages the communication cost of the system by controlling the frequency of updates between the participants and the server. Simply put, the increase of $I$ will decrease the frequency of updates and thus reduce the communication cost. But it will also depress the benefits from collaborative learning at the same time since the `collaboration' decreases. We will evaluate the effect of $I$ through extensive experiments in the next section.

\emph{Scalability.} In the above discussion, we focus on an MLP model with one hidden layer for regression analysis. Based on the above calculations, it is easy to stack more hidden layers with \textsf{ReLU} function into the model to address more complicated problems. For example, if an additional hidden layer with $b'$ neurons is added, the only change of Eq.~(\ref{equ:mlp-loss}) is adding the layer matrix multiplication and activation function in the exponential. Because the output of the previous layer is bounded to $[-1, 1]$, the sensitivity of the loss function will slightly change to $\frac{1}{2}b' + \frac{1}{8}b'^2$.

\begin{table}[tt]
\centering
\begin{tabular}{c|c ccc}
    \hline
     {Dataset}  & Baseline & Clipped ReLU & Truncated Polynomial   \\
    \hline
  US &  0.11 & 0.11 & 0.11 \\
  MNIST & 96.48 & 95.43 & 95.94 \\
  SVHN & 90.81 & 90.65 & 90.35 \\
  \hline
\end{tabular}
\caption{Accuracy of truncated functions}
\label{tab:truncation}
\end{table}

Moreover, the functional mechanism can also be applied to other types of loss functions (e.g., huber-loss function), other activation functions (e.g., hyperbolic tangent), and other types of networks (e.g., Auto-Encoder or RNN) with certain adaptations, which are beyond the scope of this paper.

\subsection{Security Analysis}\label{sec:analysis}
Let the privacy budget used in participant sampling and objective function perturbation be $\epsilon_1$ and $\epsilon_2$ respectively.
Since the training procedure on each participant strictly follows FM, the parameters computed from the perturbed objective function satisfy $\epsilon$-differential privacy in each training iteration. Let $S_i$ be the training batch of an iteration. Since all batches are disjoint from each other in a training \emph{epoch} (e.g., $S_i$ and $S_{i-1}$ contain different tuples sampled randomly from the training data), where an epoch is one full training process that consists of several iterations covering the whole training data, we can conclude that the training process at each participant ensures $\epsilon_2$-differential privacy in each epoch according to Theorem~\ref{thm:parallel}.

Recall that the sampling procedure in Algorithm~\ref{alg:server} also ensures $\epsilon_1$-differential privacy at each sampling step. Since it can be seen that each step of sampling protects the privacy of the partial training data, we can conclude that the sampling procedure in Algorithm~\ref{alg:server} satisfies $\epsilon_1$-differential privacy in each epoch.

Note that the two procedures above address two different privacy concerns respectively. In case of passive adversaries, the procedures executed by each participant aim to protect the privacy of the training data, focusing on each single record of the training data, while the procedures run by the server aim to protect the privacy of the data quality, which takes all the corresponding records as a whole. Therefore, we can finally conclude that SecProbe satisfies $\max(\epsilon_1, \epsilon_2)$-differential privacy in each training epoch.
%Therefore, we can conclude that SecProbe satisfies ($\epsilon_1$+$\epsilon_2$)-differential privacy in each training epoch.}}

We further consider the effects of different behaviors by adversaries. Firstly, if a set of participants are \emph{curious-but-honest}, they may try to infer some information from others. But obviously, using differential privacy can prevent this leakage. If some participants are \emph{malicious}, there may be two kinds of behaviors: 1) sending fake parameters to the server; 2) stealing the parameters from the communication process directly. For the first malicious behavior, thanks to the exponential mechanism that we have introduced, it is almost impossible for the fake parameters to significantly affect the model, thus there is no negative effect on the training process, and the adversaries cannot infer the data quality of other participants. For the second one, the adversary may eavesdrop on channels between honest participants and the server. Some effective cryptography tools can be used to encrypt the parameters (e.g., AES) and verify the received data (e.g., SHA-256) to ensure communication security. Therefore, our scheme is robust and secure when facing malicious participants.

Furthermore, in this paper, we assume the server is honest-but-curious and non-colluding, i.e., the server can know the data quality of each participant after uploading, but it will not reveal the qualities or collude with some participants. As discussed in~\cite{shokri2015privacy}, anonymous authentication and communication techniques can help to relax this assumption by preventing the server from linking uploaded parameters with the participants. At the beginning of the protocol, a participant can use the protocol in~\cite{camenisch2006win} to obtain enough e-tokens (e.g., in our experimental settings, 500 is sufficient to make the learning process converge), and in each round of communication the participant can use an e-token to authenticate his identity and upload the parameters. Then, the approach in~\cite{wolinsky2012dissent} can be used to prevent tracking the identity while downloading the updated model. Obviously, the anonymity only relies on the security of the adopted anonymous communication protocol, the design of which is outside the scope of this paper.

However, even if the identities of participants are protected by the anonymity techniques, the use of differential privacy is still essential. Considering an extreme case that if only one participant is honest, and all the other $N-1$ participants are collusive. Obviously, the parameters of the honest one are easy to be inferred. This is because the aggregation is only a simple average, and some sensitive information in the local dataset of the honest participant would be leaked from the parameters, even if the others do not know which party owns these information. In this scenario, differential privacy will work as differential privacy can help protect the record privacy even if the adversaries know all the other records in the dataset.

\section{Experimental Evaluation} \label{sec:evaluation}
In this section, we evaluate the performance of SecProbe on a real-world dataset.
%\sout{We implement SecProbe on TensorFlow ~\mbox{\cite{Tensor}}, a popular deep-learning library developed by Google, and }
All experiments are conducted on a machine with an Intel E5-2660-v2 CPU, a Quadro K5200 GPU and 128GB RAM, running on Ubuntu 16.04.

\subsection{Datasets}
For the regression task, we use the dataset from \emph{Integrated Public Use Microdata Series}~\cite{Minnes}, named US, which contains 600,000 census records collected in US. There are 15 attributes in the dataset, namely, \emph{Sex, Age, Race, Education, Filed of Degree, Marital Status, Family Size, Number of Children, Hours Work per Week, Ownership of Dwelling, Number of Children, Number of Rooms, Private Health Insurance, Living Difficulty and Annual Income.} Among all these attributes, there are 6 attributes that are categorical, including \emph{Race, Education, Field of Degree, Marital Status, Private Health Insurance and Living Difficulty.} For an attribute that can only be two possible values (e.g., male and female for sex), we set it to be 0 or 1. For the remainings, we follow the common practice in machine learning to transform these attributes by \emph{one-hot encoding}. We then normalize the other numeric attributes into the scope of $[0,1]$. Specifically, for the \emph{Annual Income}, we apply log transformation before normalization to obtain a relatively stable distribution~\cite{ZouNWLW18}. After these transformations, our dataset now has 20 attributes.

We randomly sample 90,000 records to construct the test dataset, and 10,000 records to form the auxiliary validation dataset on the server. The remaining 500,000 records are randomly divided into $N$ parts, where $N$ is the number of participants. The data is already shuffled before training.

We focus on a regression task predicting the value of \emph{Annual Income} by using the other attributes as the input. The accuracy of the model is measured by \emph{mean relative error} (MRE),
\begin{equation}
 \textbf{MRE} = \frac{1}{n}\sum_{i=1}^n \frac{|z_i-y_i|}{y_i},
\end{equation}
where $y_i$ is the real value, $z_i$ is the predicted value produced by the network, and $n$ is the number of tuples in the test dataset.

For the binomial classification task, like~\cite{shokri2015privacy}, we use the MNIST and SVHN datasets as benchmarks. The MNIST dataset consists of 28x28 images of handwritten digits with 60,000 training samples and 10,000 test samples. The SVHN dataset consists of 32x32 images of house numbers with 73,527 training samples, 26,032 test samples and 10 classes. We use the accuracy of classification to evaluate the performance of these two models.

%\begin{figure*}[t]
%  \centering
%  \subfigure[$N$ = 30 ]{
%  \includegraphics[width = 0.25\textwidth]{./pic/change_M_30.pdf}}
%  \subfigure[$N$ = 60 ]{
%  \includegraphics[width = 0.25\textwidth]{./pic/change_M_60.pdf}}
%  \subfigure[$N$ = 100 ]{
%  \includegraphics[width = 0.25\textwidth]{./pic/change_M_100.pdf}}
%    \vspace{-2mm}
%  \caption{The effect of number of participants ($M$) utilized per round on the system performance.}\label{fig:M-participants}
%    \vspace{-5mm}
%\end{figure*}

\subsection{Experimental Setup}
We use the popular neural network architectures: multi-layer perceptron (MLP) with three fully-connected layers. For the regression task, the activation functions of hidden layer and output layer are ReLU and sigmoid function respectively. The number of neurons in the hidden layer is 80. We use SGD as the learning algorithm. The learning rate and the mini-batch size are set to be 0.01 and 128, respectively. The weights of the models are randomly initialized by normal distribution (with mean 0 and standard deviation 1).

For both MNIST and SVHN datasets, we use a CNN with two convolutional layers, two max-pooling layers, and one hidden layer with 128 neurons. The activation of the output layer is the sigmoid function. The other hyperparameters are the same as the regression task.

%Due to the absence of existing schemes on solving our target problem,
Since the approaches proposed in~\cite{phan2016differential} and~\cite{abadi2016deep} are not specially designed for collaborative learning, we mainly compare all results with DSSGD in~\cite{shokri2015privacy} and~\cite{geyer2017differentially}, and two baseline approaches. The first is the \emph{centralized training on the entire dataset}, which is the basic approach that does not consider privacy concerns and ought to have the best performance of the model accuracy. The second is \emph{stand-alone training}, which trains solely on the local dataset without collaboration. We call these two baselines \emph{Centralized} and \emph{Stand-alone}, respectively. For simplicity, we abbreviate the method in~\cite{geyer2017differentially} as CSDP in the rest of this paper. All schemes are implemented on TensorFlow. For SecProbe, we set the privacy budgets used in participants' sampling and perturbing objective functions to be the same. We fine tune the parameters in DSSGD according to~\cite{shokri2015privacy} and use the settings with the best performance (the parameter download ratio $\theta_d=1$, gradient bound $\gamma=0.001$, gradient selecting threshold $\tau=0.0001$).

To simulate the unreliable participants, we randomly choose half of the participants and replace $P$ fraction of their data with random noise in the range [0,1]. We vary $P$ to evaluate the robustness of SecProbe against unreliable participants.

\begin{figure*}[t!]
  \begin{minipage}{0.95\textwidth}

  \subfigure[N = 30]{
  \includegraphics[width = 0.34\columnwidth]{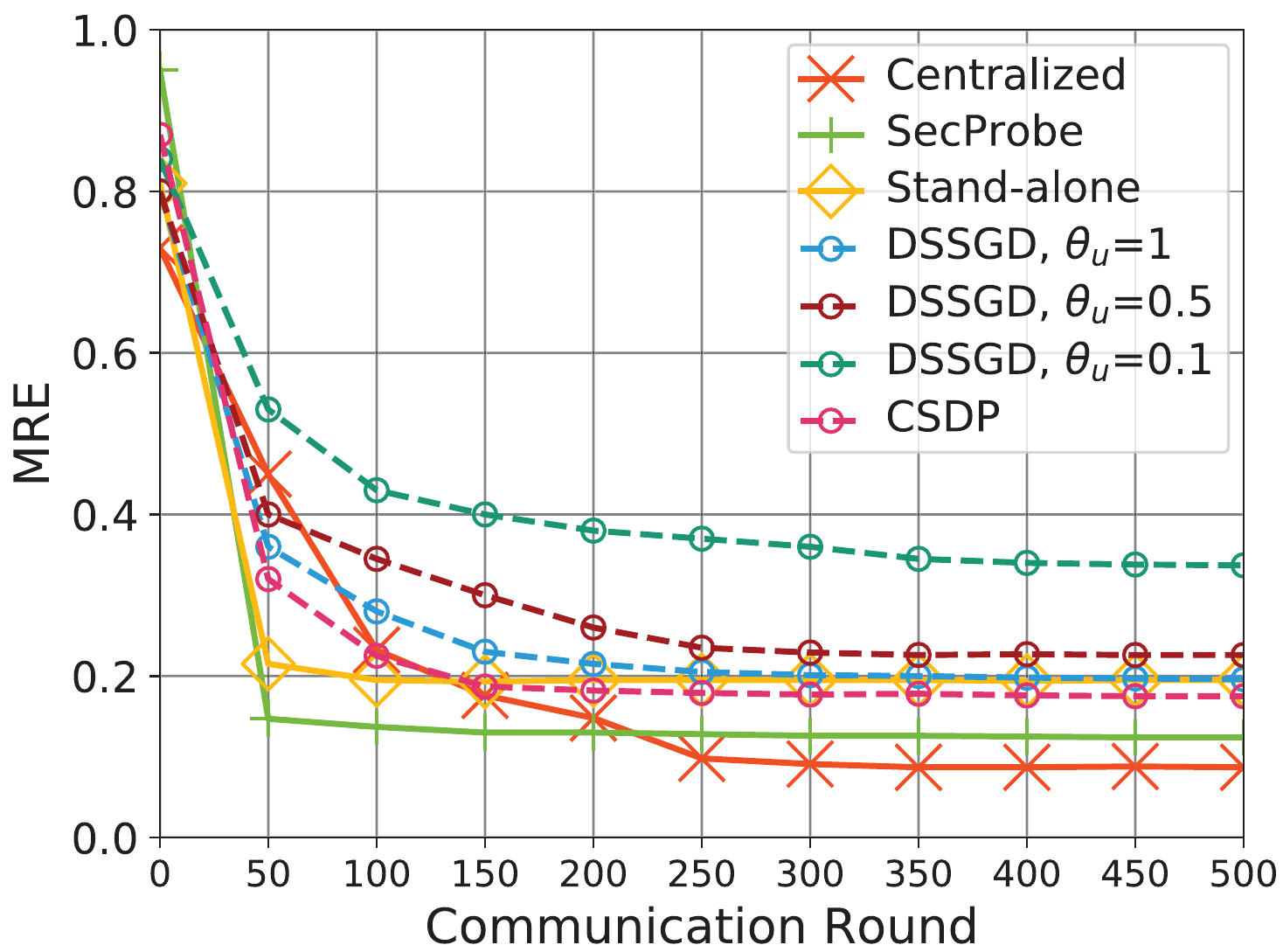}}
  \subfigure[N = 60]{
  \includegraphics[width = 0.34\columnwidth]{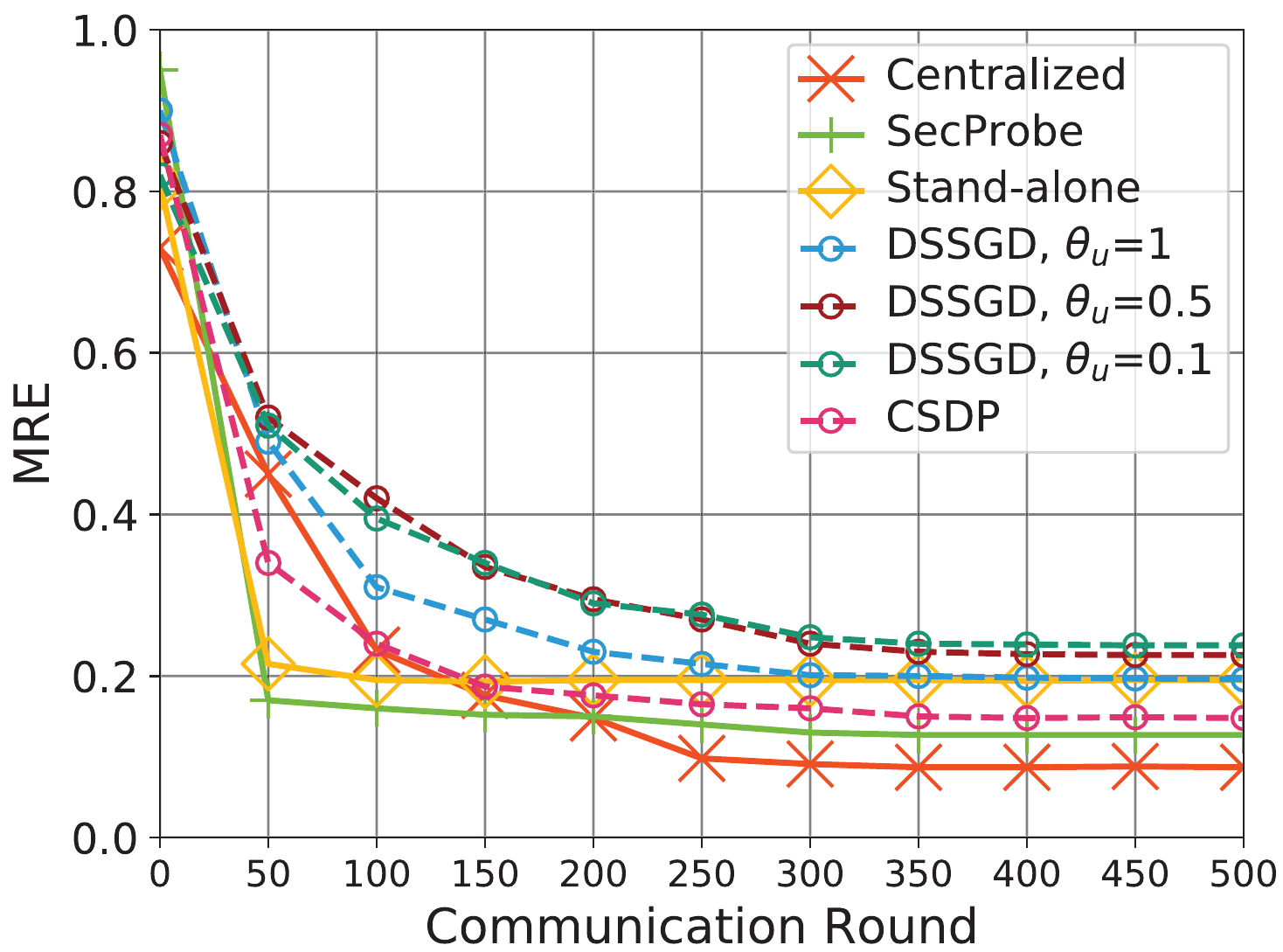}}
  \subfigure[N = 100]{
  \includegraphics[width = 0.34\columnwidth]{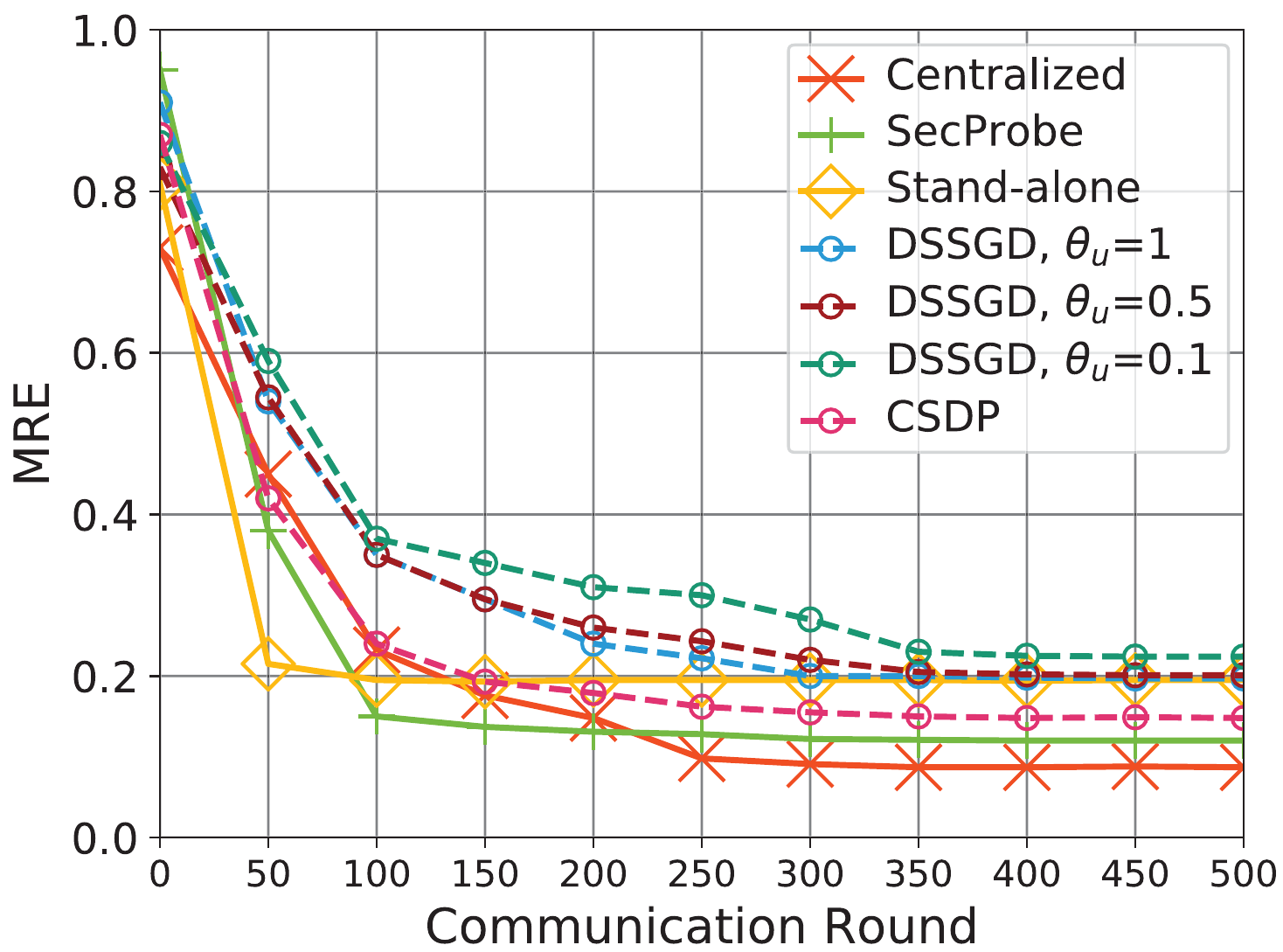}}
  \caption{A comparison of training convergences of all schemes for the regression task}\label{fig:N-regression}
\end{minipage}
\end{figure*}

\begin{figure*}[t!]
  \begin{minipage}{0.95\textwidth}
  \subfigure[N = 30]{
  \includegraphics[width = 0.34\columnwidth]{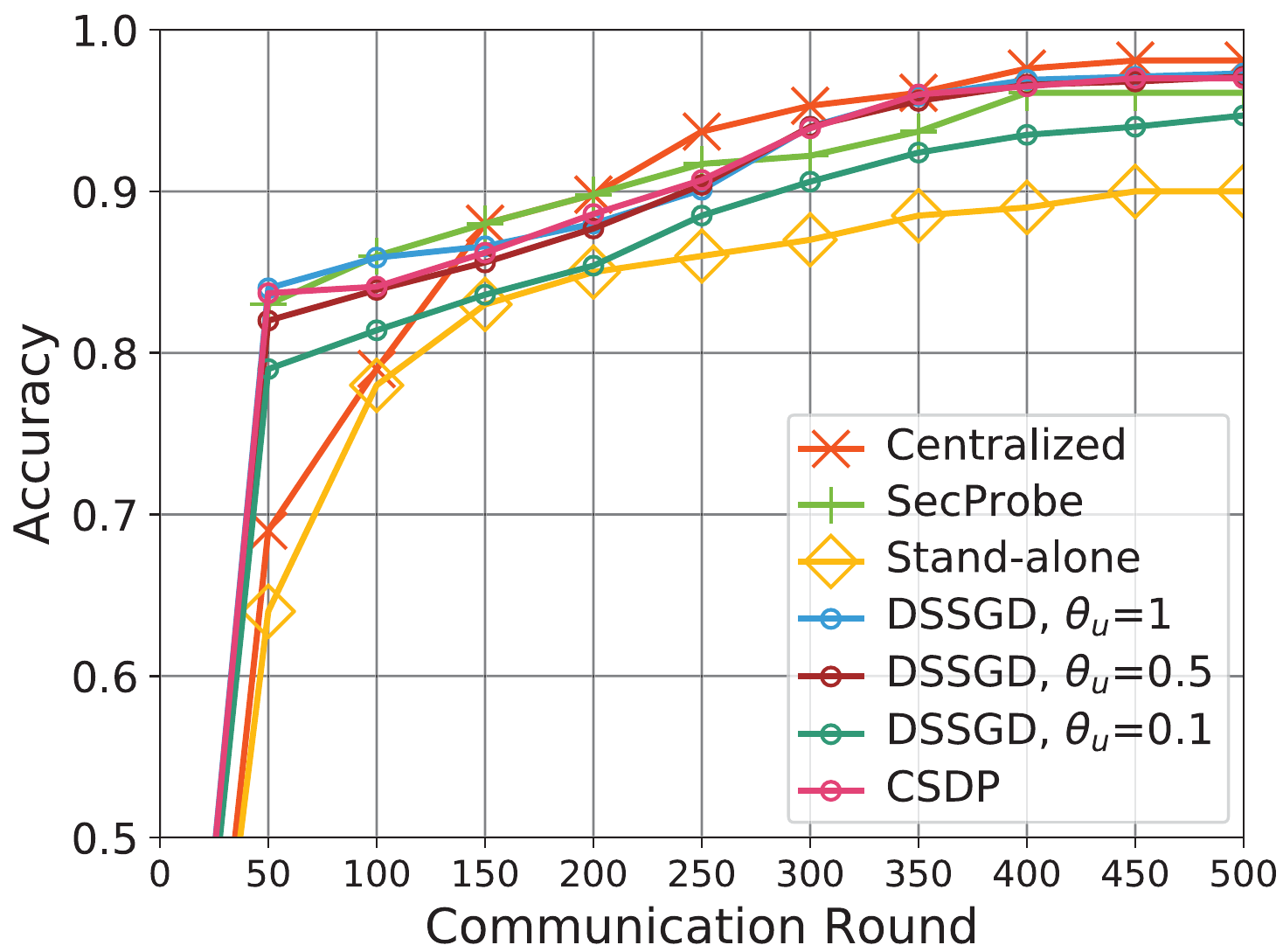}}
  \subfigure[N = 60]{
  \includegraphics[width = 0.34\columnwidth]{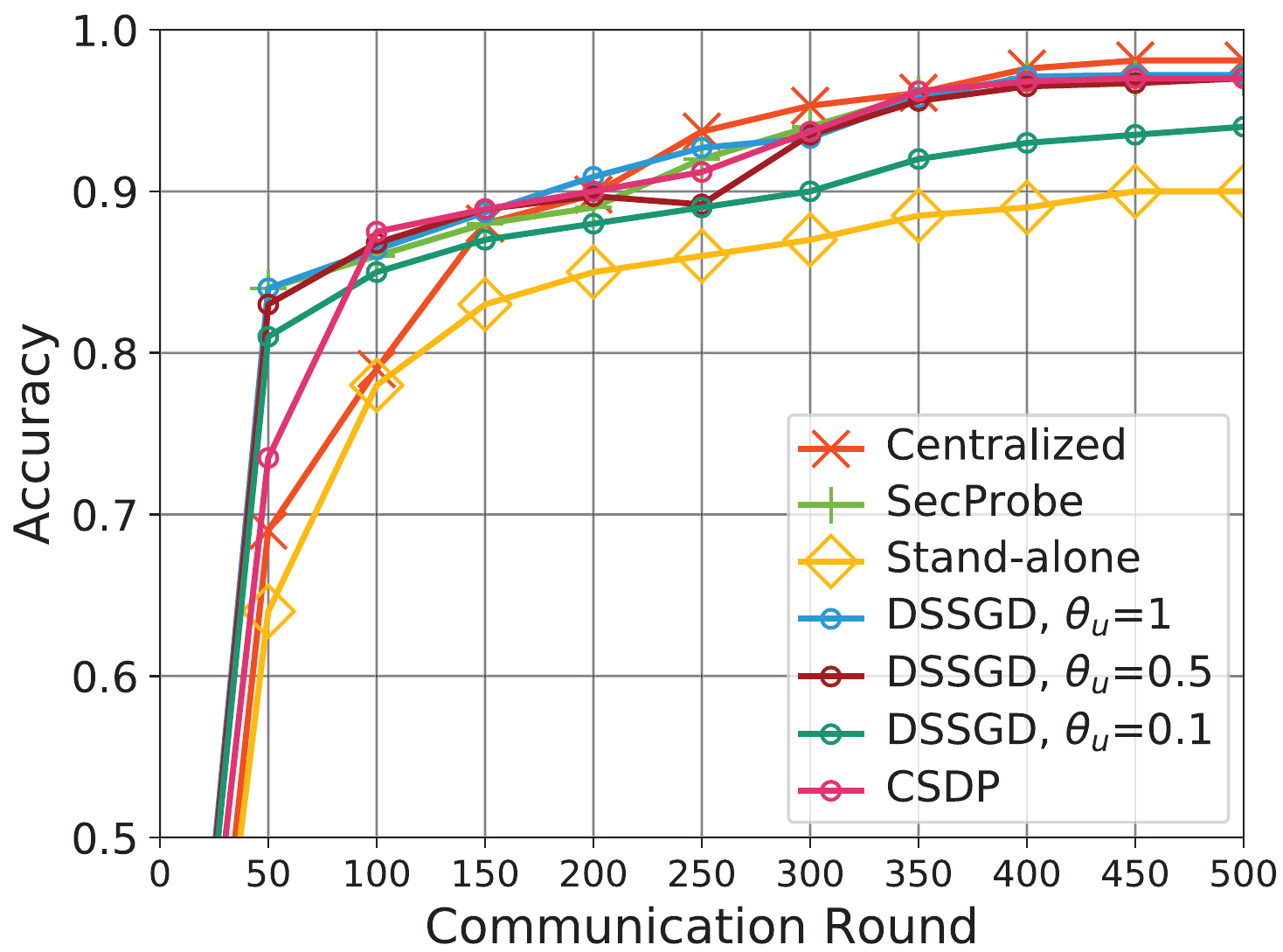}}
  \subfigure[N = 100]{
  \includegraphics[width = 0.34\columnwidth]{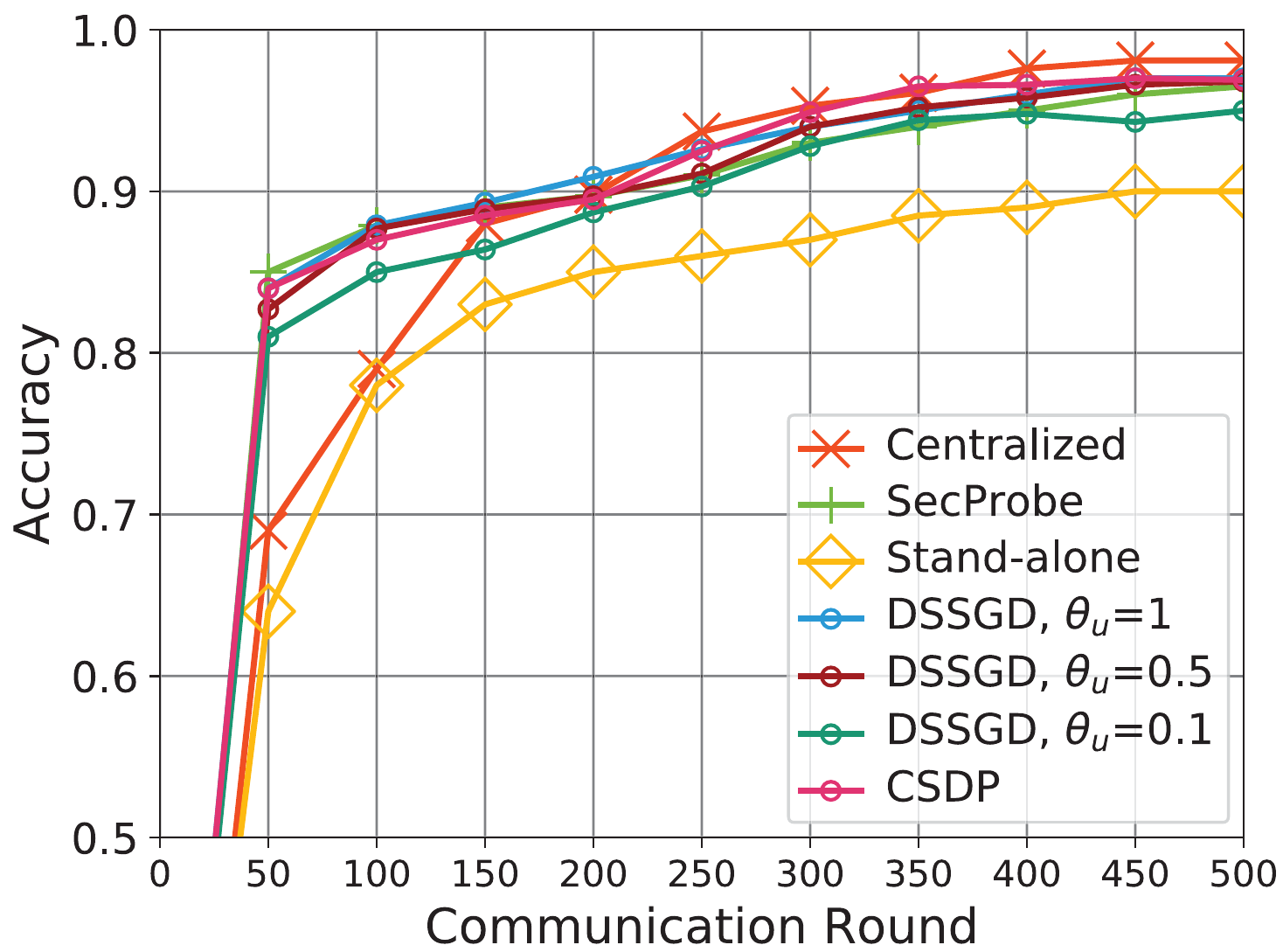}}
  \caption{Training convergences of all schemes on MNIST}\label{fig:N-classification1}
\end{minipage}
\end{figure*}

\begin{figure*}[t!]
  \begin{minipage}{0.95\textwidth}
  \subfigure[N = 30]{
  \includegraphics[width = 0.34\columnwidth]{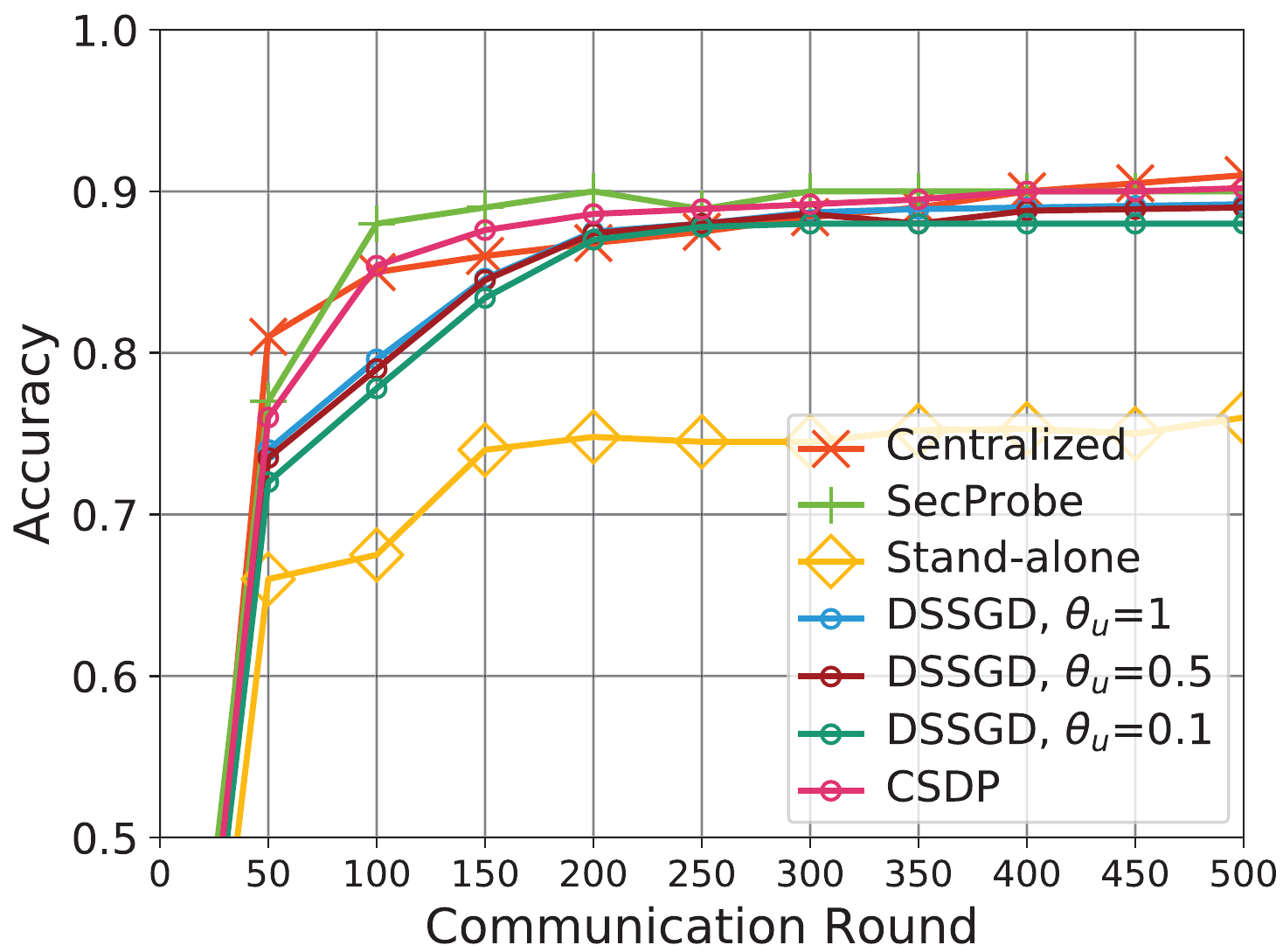}}
  \subfigure[N = 60]{
  \includegraphics[width = 0.34\columnwidth]{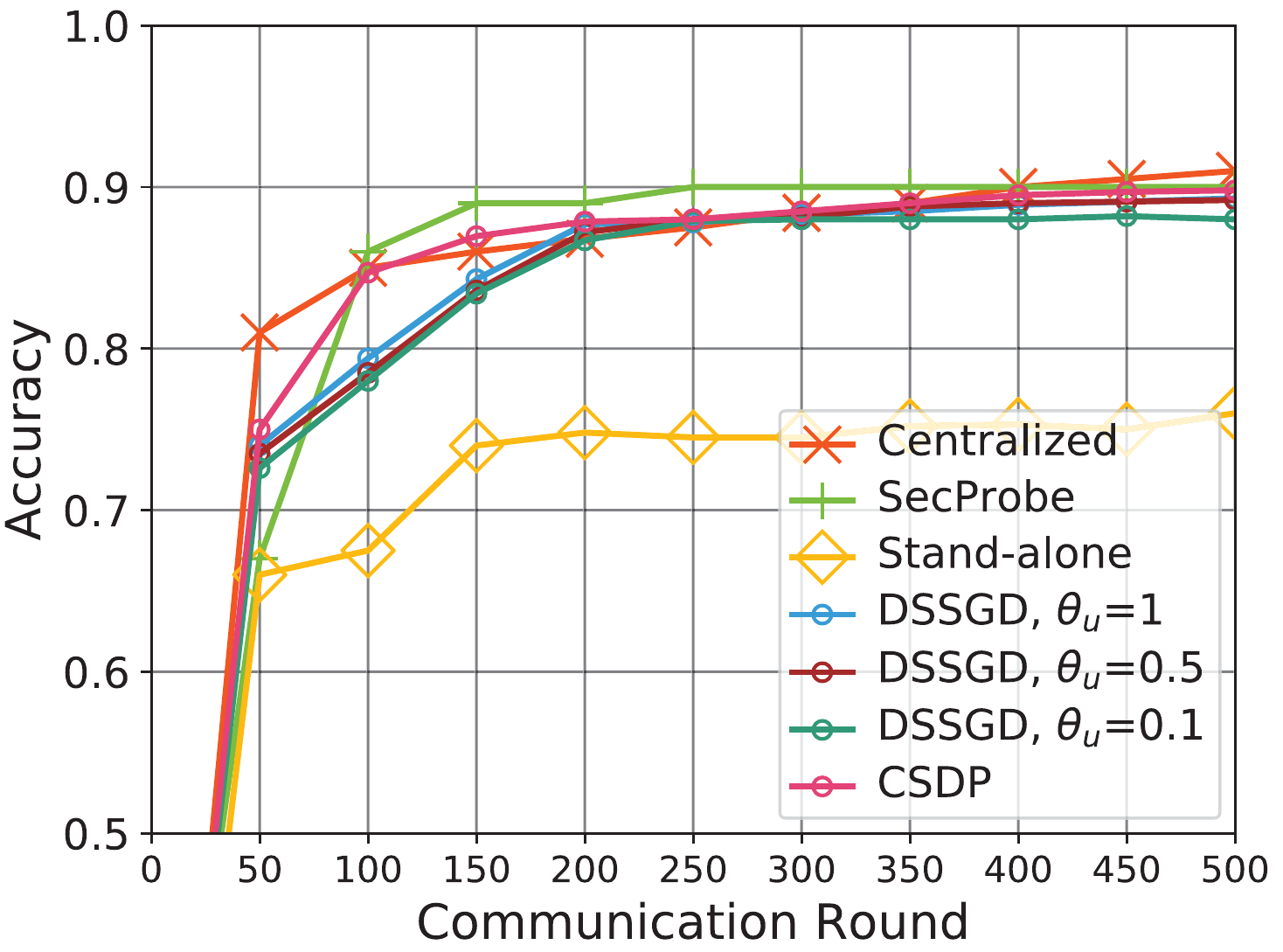}}
  \subfigure[N = 100]{
  \includegraphics[width = 0.34\columnwidth]{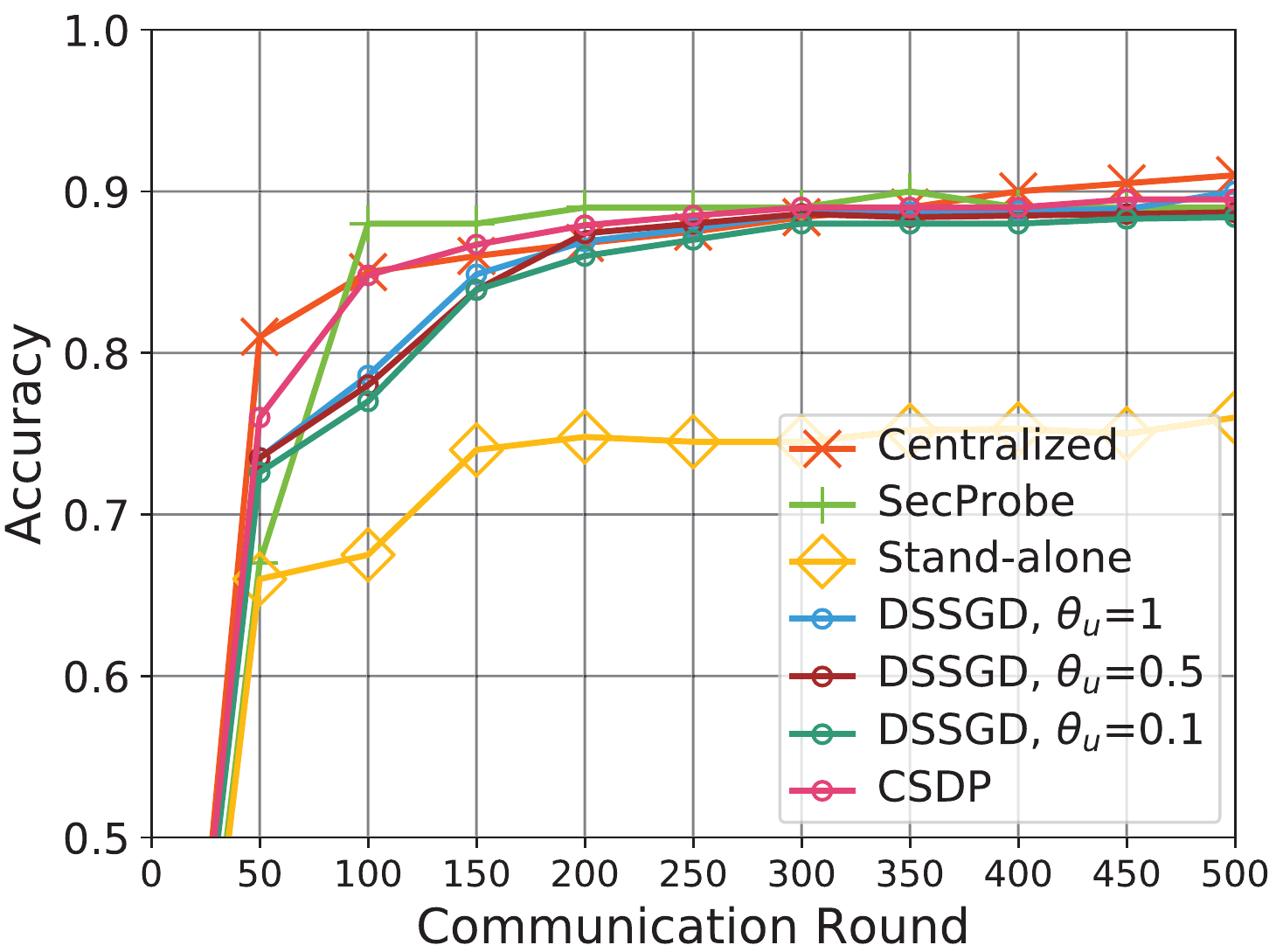}}\vspace{-3mm}
  \caption{Training convergences of all schemes on SVHN}\label{fig:N-classification2}
\end{minipage}
\end{figure*}

\begin{table}
\centering
%\tiny
\arrayrulewidth1.0pt
\begin{tabular}
{{|>{\columncolor[gray]{.8}}c
|c|c|c|c|}}
    \hline
    \rowcolor[gray]{.8}
    $I$ & 20 & 50 & 100 & 1000  \\
    \hline
  Num. of communication rounds & 443 & 190 & 152 & 193 \\
  \hline
\end{tabular}
\caption{The effect of the number of iterations $I$ which controls the frequency of updates. Each entry in the table gives the necessary number of communication rounds to achieve MRE 0.15 ($N$ = 60, $M$ = 30, $K$ = 30, $P$ = 0, and $\epsilon$ = 1).}\label{tab:iteration}
%  \vspace{-2mm}
\end{table}

\subsection{Results}

\begin{figure*}
  \begin{minipage}{0.95\textwidth}

  \subfigure[N = 30]{
  \includegraphics[width = 0.34\columnwidth]{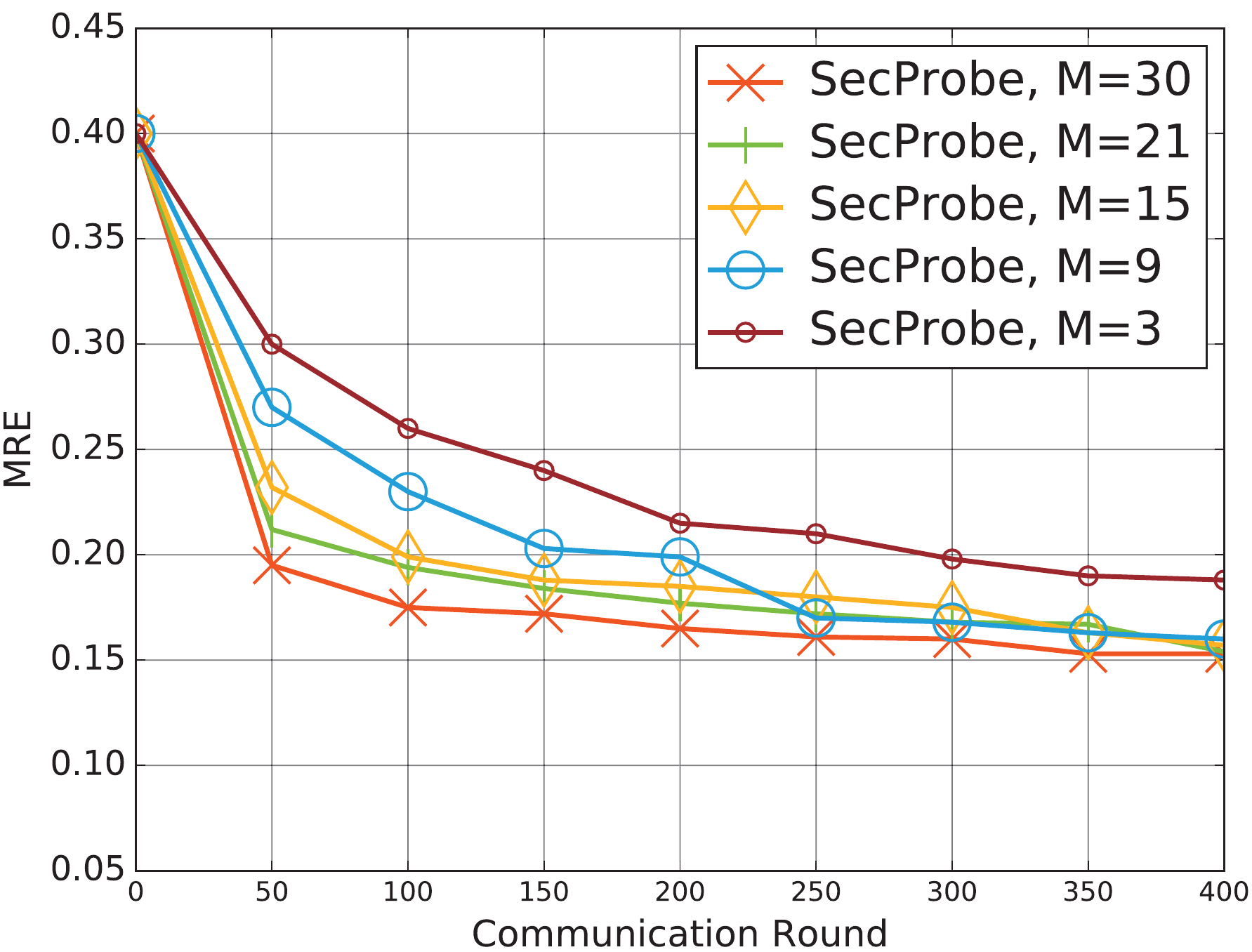}}
  \subfigure[N = 60]{
  \includegraphics[width = 0.34\columnwidth]{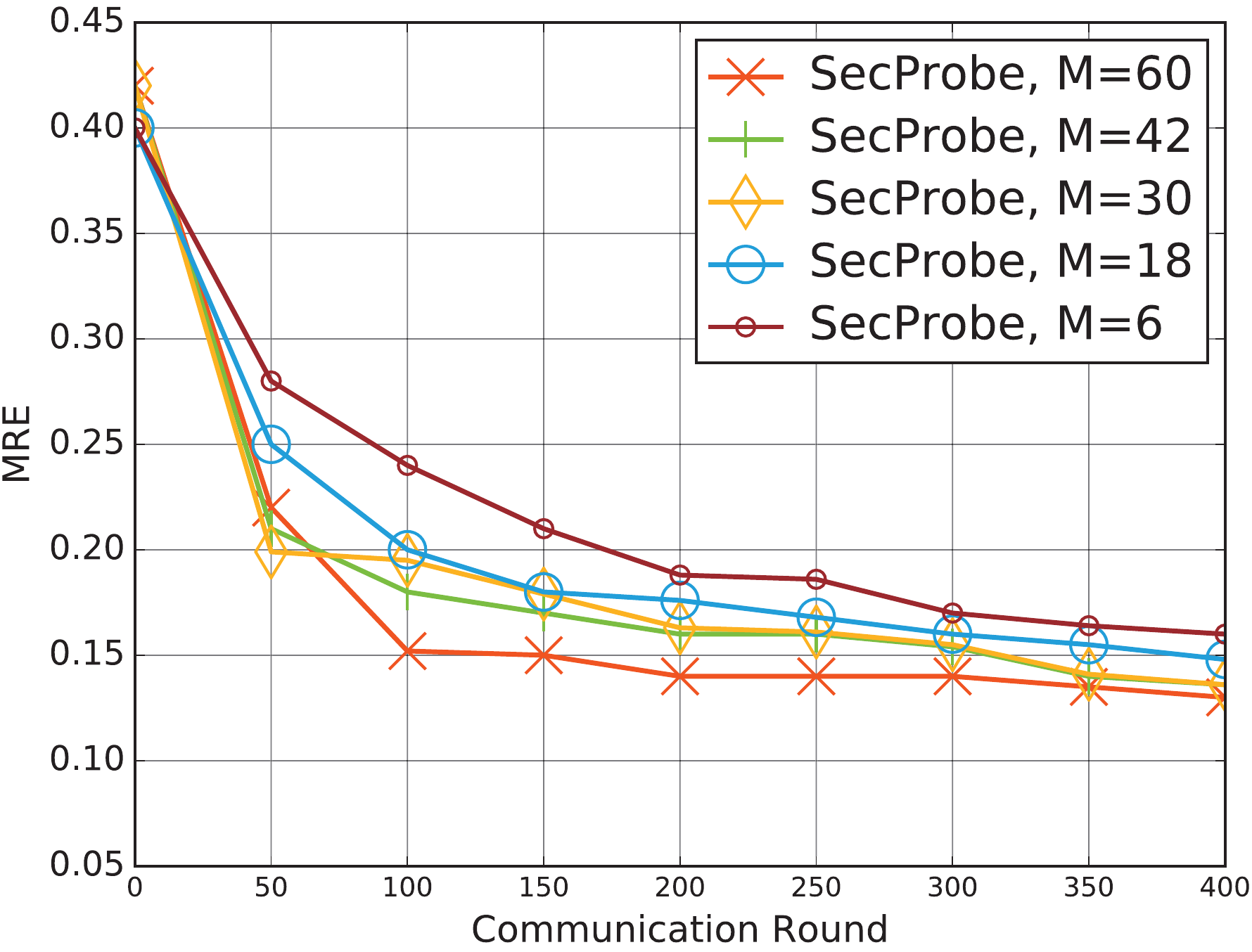}}
  \subfigure[N = 100]{
  \includegraphics[width = 0.34\columnwidth]{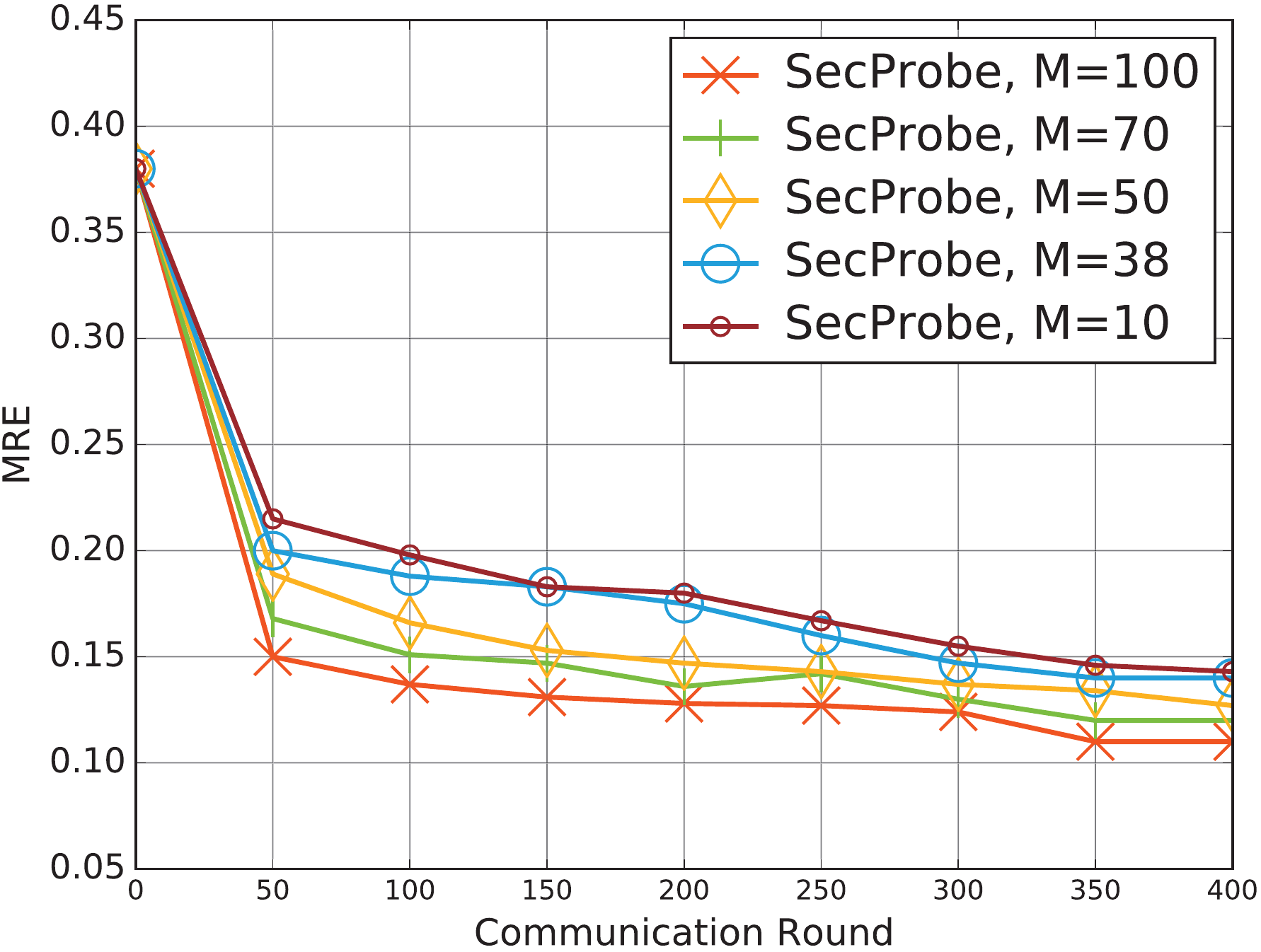}}
  \caption{The effect of the number of participants ($M$) selected per round on the system performance}\label{fig:M-participants}
\end{minipage}
\end{figure*}

\begin{figure*}
  \begin{minipage}{0.95\textwidth}

  \subfigure[N = 30]{
  \includegraphics[width = 0.34\columnwidth]{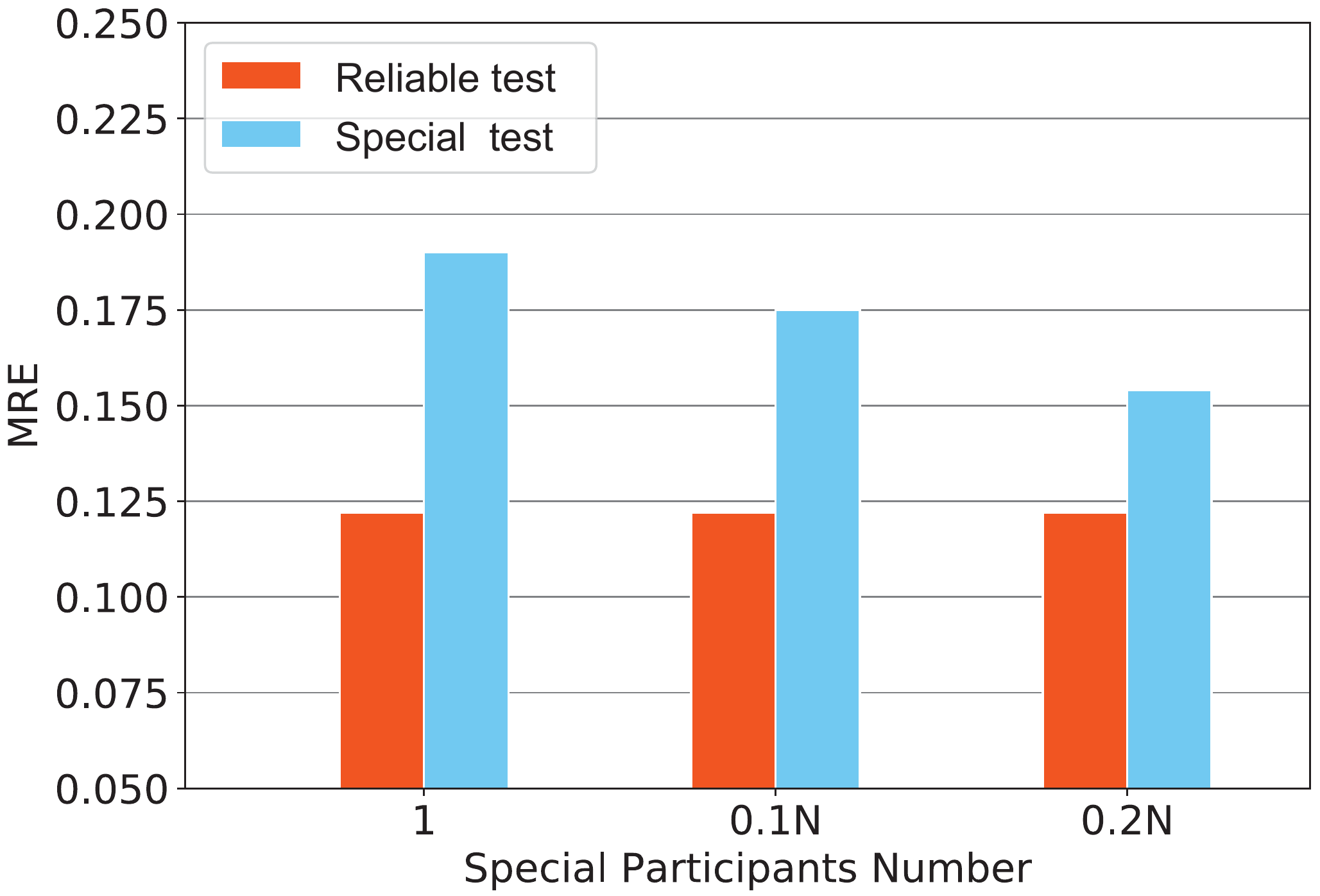}}
  \subfigure[N = 60]{
  \includegraphics[width = 0.34\columnwidth]{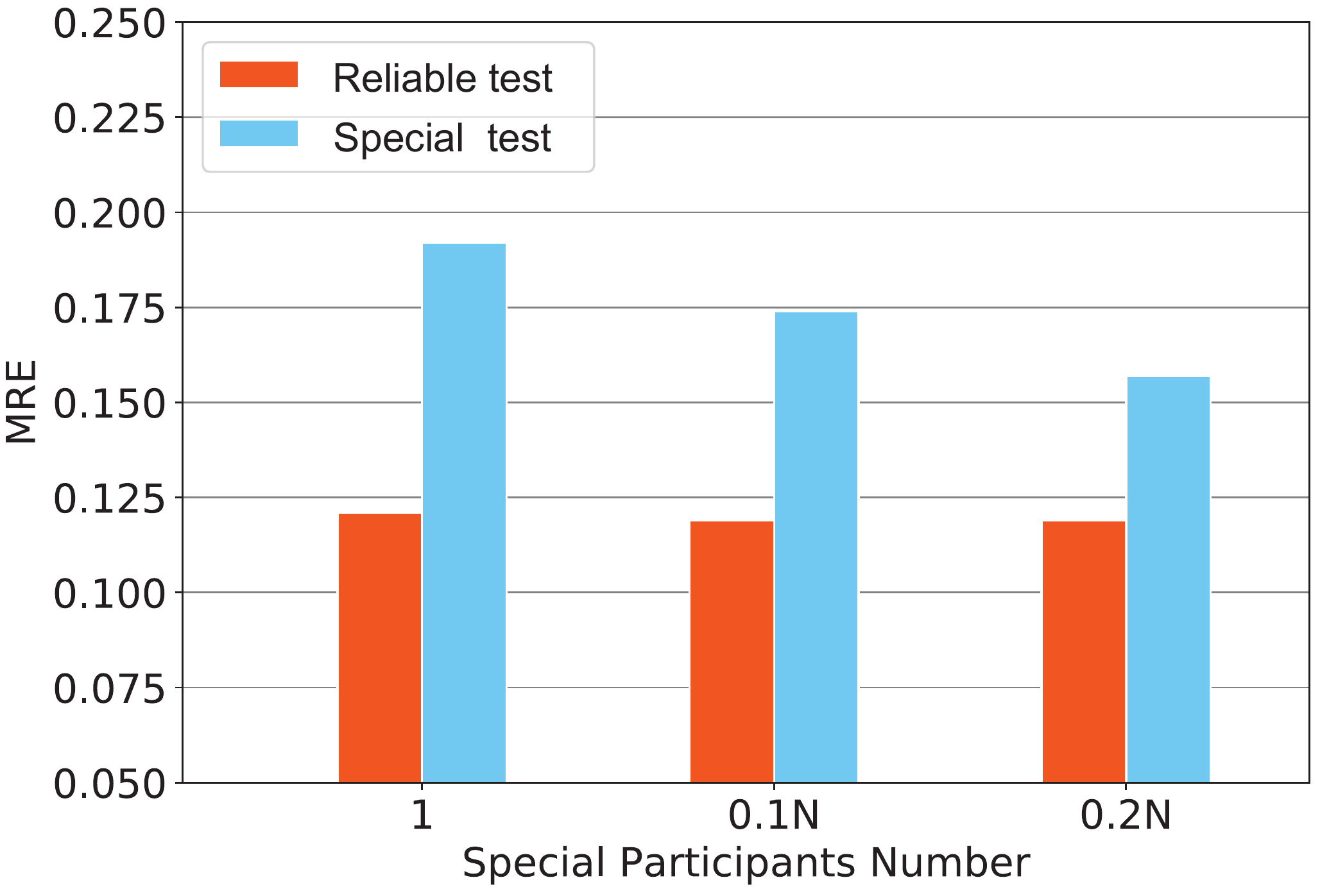}}
  \subfigure[N = 100]{
  \includegraphics[width = 0.34\columnwidth]{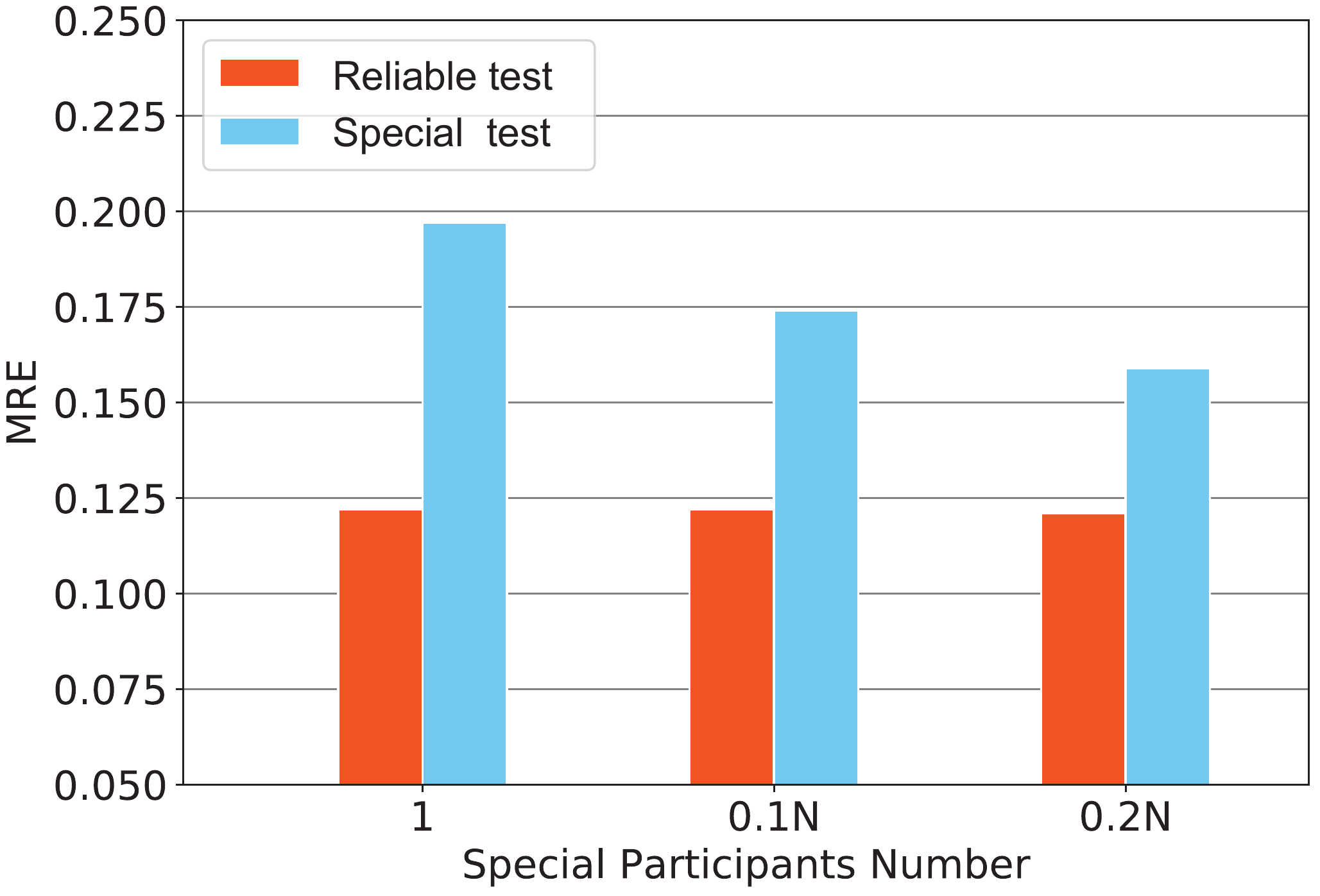}}
  \caption{The influence of the number of participants ($M$) selected per round on the system performance}\label{fig:specify}
\end{minipage}
\end{figure*}

\begin{figure*}
  \begin{minipage}{0.95\textwidth}
  \subfigure[R = 30\%]{
  \includegraphics[width = 0.34\columnwidth]{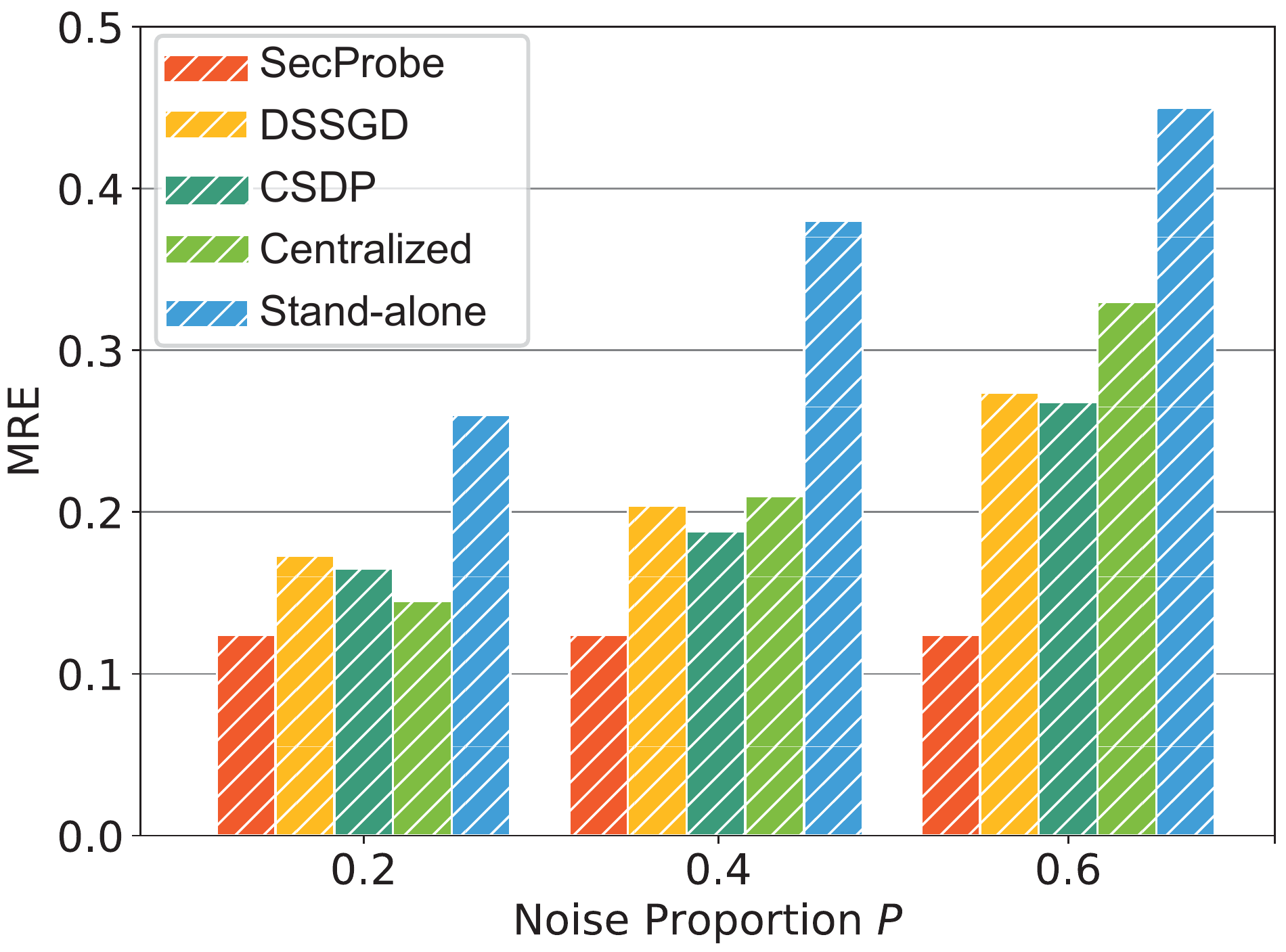}}
  \subfigure[R = 50\%]{
  \includegraphics[width = 0.34\columnwidth]{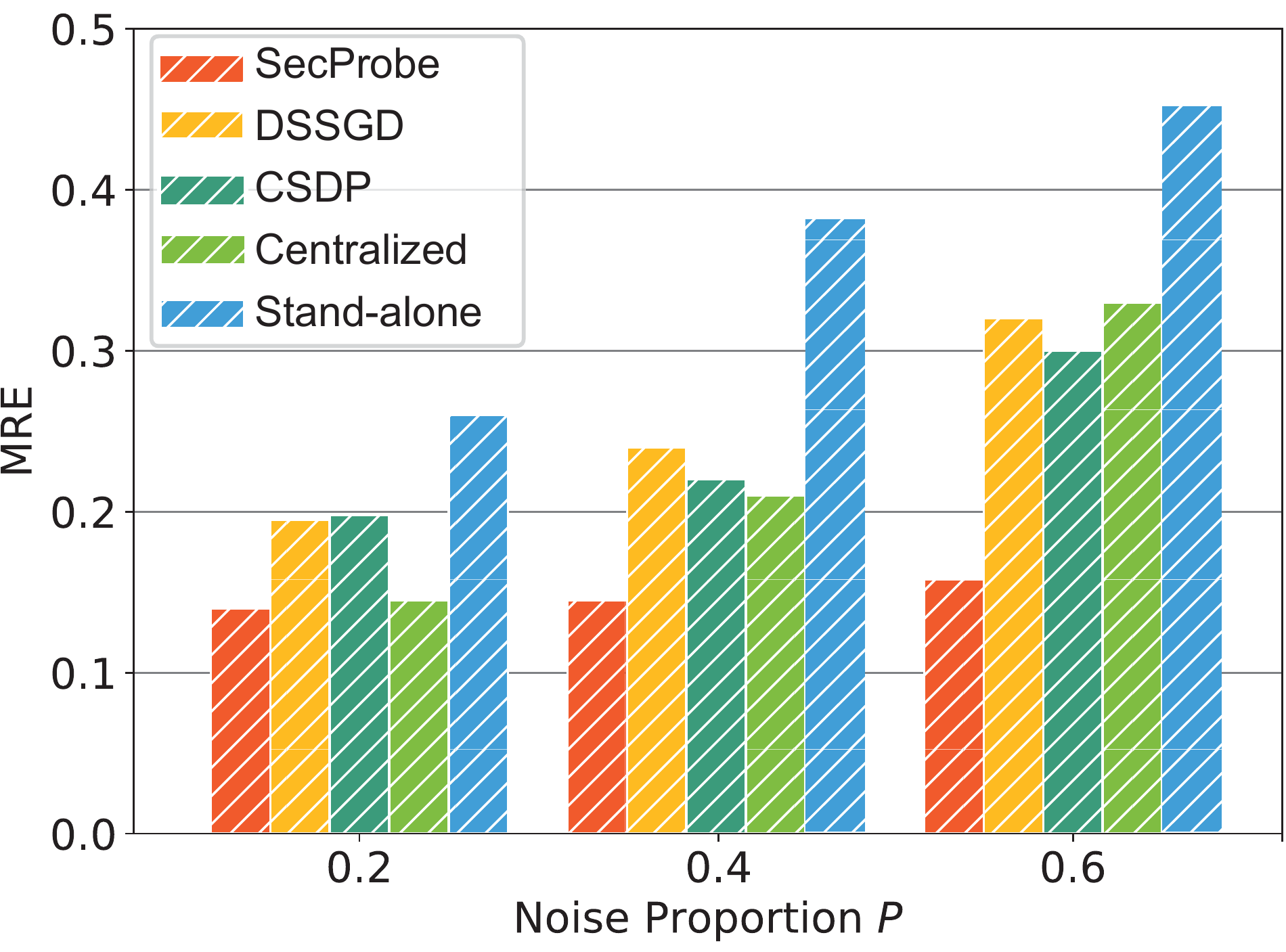}}
  \subfigure[R = 70\%]{
  \includegraphics[width = 0.34\columnwidth]{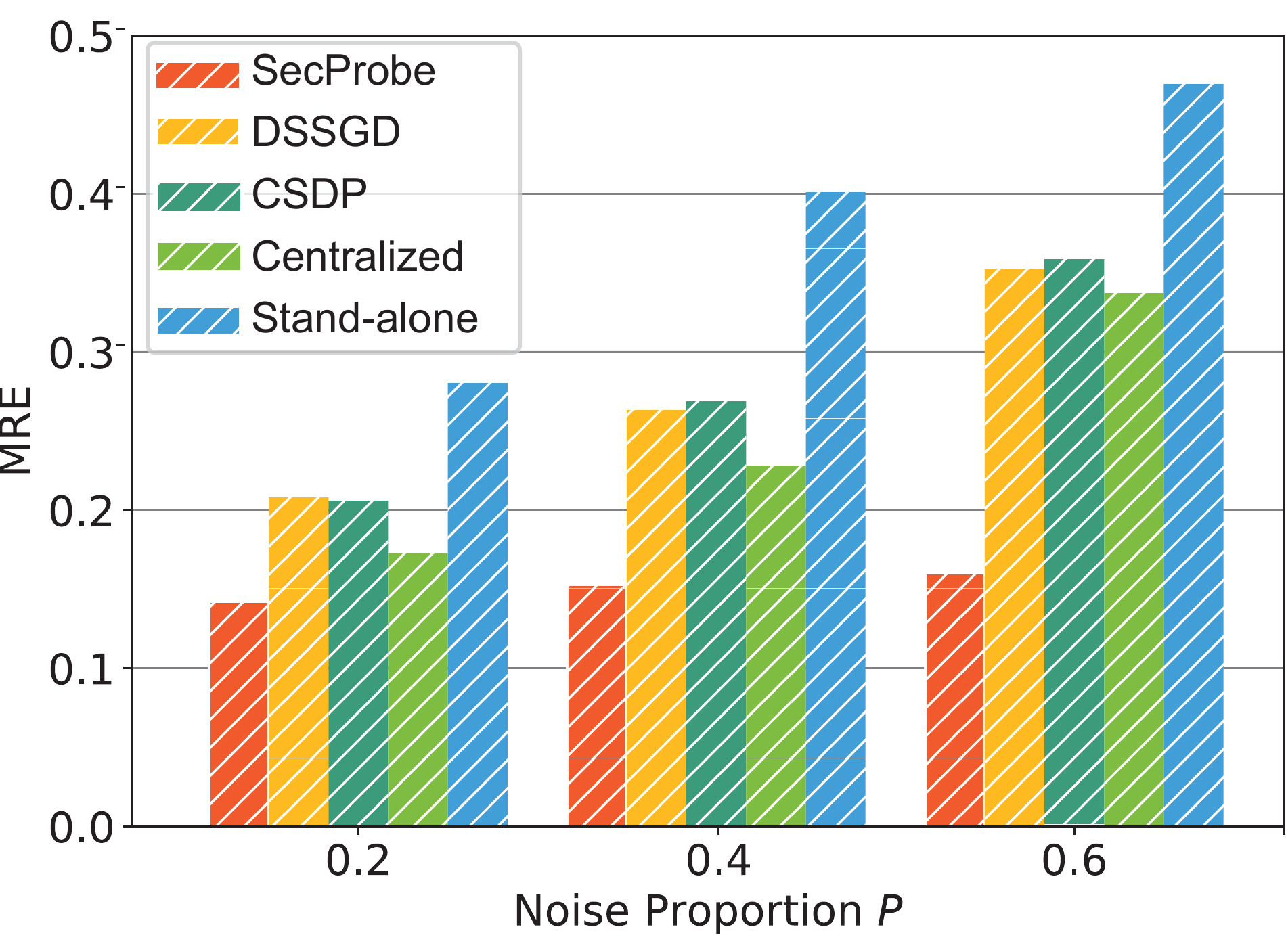}}
  \caption{The robustness of SecProbe against unreliable participants}\label{fig:noise_bar}
\end{minipage}
\end{figure*}

\textbf{The effect of I}.
Table~\ref{tab:iteration} shows the effect of parameter $I$.
%It manages the communication loads of the system by controlling the frequency of updates between
%the participants and the server.
The results show that a larger $I$ will speed up the training convergence by increasing the computation loads on each participant. However, a too large $I$ will slow down the training convergence since the collaboration decreases. Based on these results, we set $I$ to be 100 in the following experiments.

\textbf{Training convergence.} Figures~\ref{fig:N-regression},~\ref{fig:N-classification1} and~\ref{fig:N-classification2} show the training convergence of all schemes for regression and classification tasks respectively. We vary the number of participants $N$ in SecProbe and DSSGD, and set $M = N$, $K = M$, $\epsilon = 1$, and $P = 0$. The $y$-axis is the performance of trained model, and the $x$-axis denotes the number of communication rounds. Although \emph{Centralized} achieves the highest accuracy in all the settings, SecProbe can achieve almost the same accuracy while providing a rigorous privacy guarantee. In US and MNIST, \emph{Centralized} has lower convergence rate than the distributed schemes at the early stage. This may be because that more data samples are used for each iteration for the whole system, which can help accelerate the convergence in the early stage of training. Our scheme also has a better performance than DSSGD in terms of both convergence rate and model accuracy in the regression task, and it has almost the same performance in the classification task with more rigorous privacy guarantee. The main reason is that only perturbing the uploaded gradients is not enough to protect the training data, and DSSGD consumes too much privacy budget in perturbing all the gradient values. Moreover, the noise directly injected to each gradient independently may also make the training process unstable. In addition, on both MNIST and SVHN datasets, SecProbe achieves a comparable performance with CSDP and DSSGD in CNN when the fraction of uploaded parameters $\theta_u$ is 1.

\textbf{The effect of parallelism.} The parameter $M$ controls the number of participants that the server chooses per round, which can also be regarded as parallelism degree. We vary $M$ to be ($0.1N, 0.3N, 0.5N, 0.7N, 1.0N$) and set $K = M$, $\epsilon = 1$, and $P = 0$. Note that the total size of training dataset does not change with different values of $M$. As can be seen in Figure~\ref{fig:M-participants},
%The $x$-axis denotes the communication round.
the increase of parallelism will speed up the convergence of training, and it leads to a more accurate model while increasing the communication loads. On the contrary, the decrease of parallelism will reduce the number of accesses for each data point. Briefly speaking, the size of $M$ affects the amount of data for training in each communication round and the convergence speed. It is good to see that, when $M=0.1N$ it can still achieve relatively accurate results. Based on the above results, we choose $M = 0.5N$ for the following experiments to strike a good balance between efficiency and convergence rate.

\textbf{The effect of auxiliary validation set.} To show the effect of the auxiliary validation set on the system performance, we re-design the training set and the test set to simulate one or multiple \emph{special participants} whose hypotheses are not included at the server. More specifically, we exclude all samples whose one attribute is within a certain range from the original datasets, and put these excluded data to the special participants and a special test set. Here we choose the attribute $\emph{Age}$ due to its numeric form and its impact on the income. We remove all samples with $\emph{Age}\leq 0.25$ or $\emph{Age}\geq 0.7$ from the original datasets. We set the number of special participants as 1, 0.1N, and 0.2N respectively, and demonstrate the results in Figure~\ref{fig:specify}. We can see that the performance on the special test set is significantly worse than that of the reliable test set. With more special participants, the performance of the special test set will improve due to the increasing contribution of the special participants to the collaborative learning process. Therefore, a special participant with high-quality dataset that contains new findings can also play a role in improving the model, even if it is misjudged by the server.

Certainly, the lack of the auxiliary validation dataset which can provide good utility scores for all participants is a practical limitation. However, this can be mitigated by choosing more participants in each round, i.e., increasing the probability that special participants are chosen. It can be observed from Figure~\ref{fig:specify} that even if the special participants' models do not have good utility scores, the aggregated model can have the increased performance on the special test set while rising the proportion of special participants.

\begin{figure}[t!]
\centering
\includegraphics[width=0.70\columnwidth]{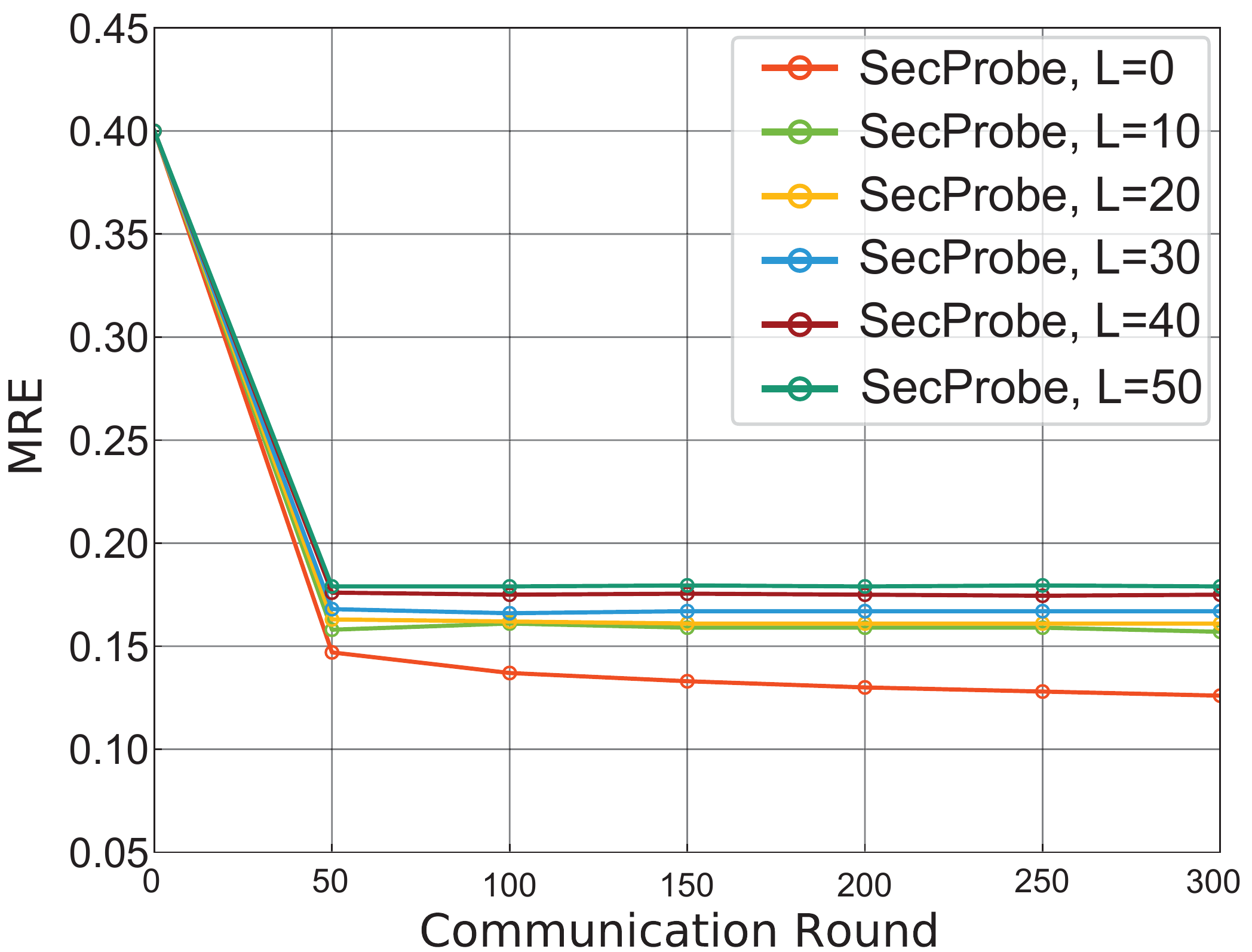}
\caption{The influence of malicious participants}\label{fig:malicious}
\end{figure}

\textbf{Robustness against unreliable participants.} Figure~\ref{fig:noise_bar} shows the results of the robustness against unreliable participants. We vary the proportion of unreliable participants $R$ as (30\%, 50\%, 70\%), and replace $P$ proportion of their data with random noise, and set $M = 0.5N$, $K = 0.5M$, and $\epsilon$ = 1. We vary the total number of participants $N$ and the proportion $P$ of the noise data, which means that one half of the participants are unreliable with $P$ fraction of their data to be random noise. We set $N = (30, 60, 100)$ and $P = (0.2, 0.4, 0.6)$. Correspondingly, for \emph{Centralized}, we set $P = (0.1, 0.2, 0.3)$. For the number of sampling participants $K$, we set it to be the half of $M$, which follows the assumption that the majority of the participants are reliable. As can be seen in Figure~\ref{fig:noise_bar}, the prediction accuracies of all the other methods decrease quickly with the increase of noise, and more unreliable participants will increase this impact, since they are all lack of the strategy dealing with unreliable participants/data. For \emph{Centralized} and \emph{Stand-alone}, they are disturbed by the unreliable data directly. Specially, for DSSGD and CSDP, they randomly sample parameters/participants to aggregate the model, under an equivalent sampling probability. As a result, the unreliable participants will significantly impact on the accuracy. Meanwhile, our approach SecProbe achieves very high accuracy which is almost the same as the performance of the case with no unreliable participants, and it is also robust against the proportion of noise. The experimental results validate the effectiveness of our scheme.
\begin{figure}[t!]
\centering
\includegraphics[width=0.70\columnwidth]{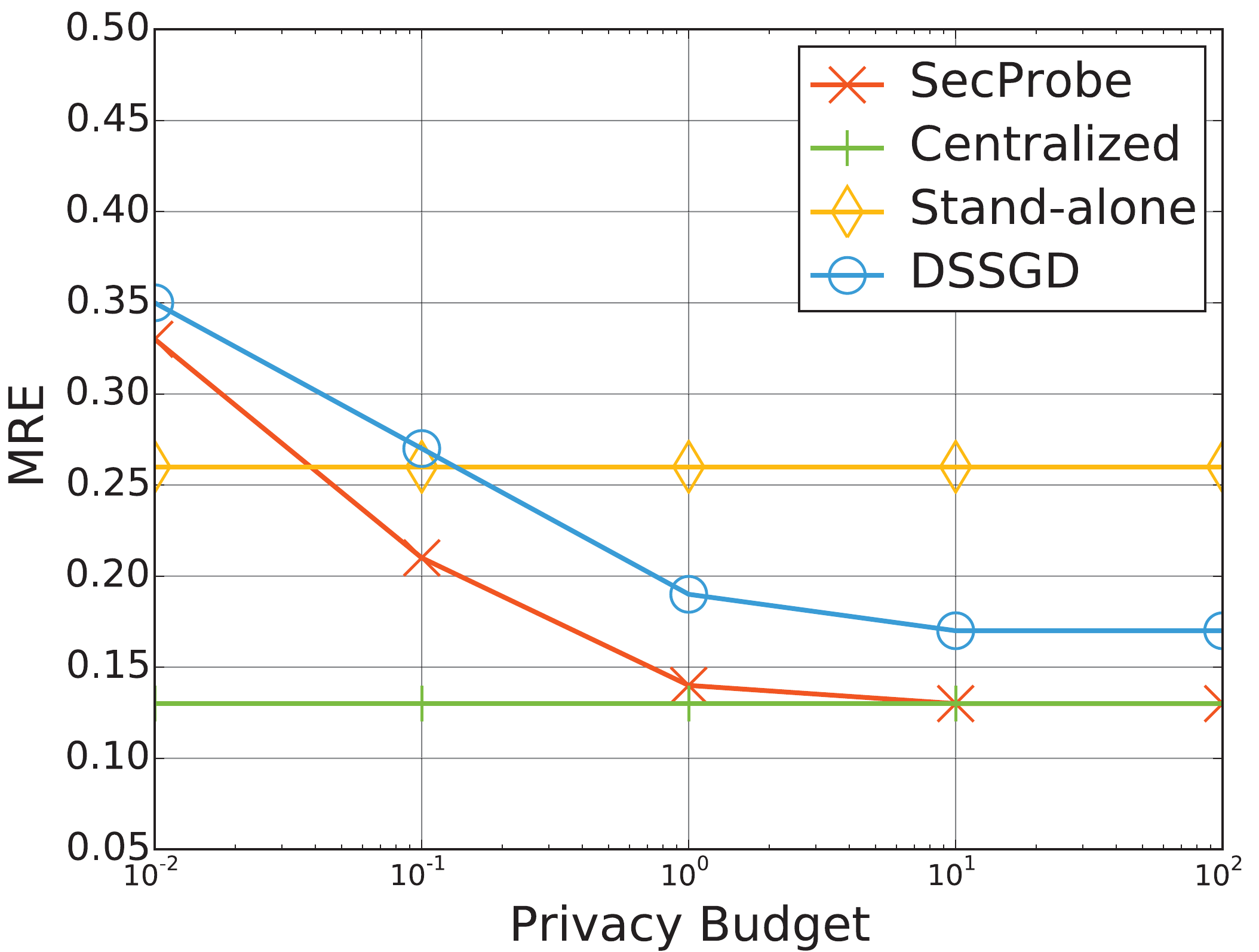}
\caption{Accuracy vs. privacy budget $\epsilon$ ($N$ = 60, $M$ = 30, $K$ = 15, and $P$ = 0.2)}\label{fig:epsilon}
\end{figure}
\textbf{The influence of malicious participants.} There may also exist some malicious participants who may destroy the usability of the aggregated model. We show the influence of malicious participants by setting $N = 60$, $M = 30$, and $K = 15$, and tuning the number of malicious participants from 10 to 50 at an interval of 10. The uploaded malicious parameters are randomly set within [0, 1], as the abnormal values, e.g., tens or hundreds, can be easily detected and dropped out. In Figure~\ref{fig:malicious}, we can see that the existence of malicious participants disturbs the training process significantly, with a fast convergence rate at the early stage of training and a sudden stop followed. The potential reason is that even if only one malicious participant is chosen, the randomly generated parameters will disturb the learning process. Although the performance of the trained model reduces gradually with the increase of malicious participants, the prediction error is still less than 0.2.

Similar to special participants, malicious participants can influence the performance owing to their contribution in aggregating the model. Therefore, it can also be mitigated by tuning the number of chosen participants in each round to exclude the malicious participants as much as possible.

\textbf{Accuracy vs. privacy}. We evaluate the effect of different values of privacy budget $\epsilon$ on the accuracy of the neural network. Figure~\ref{fig:epsilon} shows the results compared with the competitors. The $x$-axis represents the privacy budget per training epoch (an epoch contains several iterations over all training samples). It is shown that, a larger value of $\epsilon$ results in high accuracy while providing lower privacy guarantee. SecProbe achieves almost the same results with \emph{Centralized} and outperforms \emph{Stand-alone} when $\epsilon\geq 0.1$.

\section{Conclusions}\label{sec:conclusion}
In this paper, we took the first step to investigate the problem of privacy-preserving collaborative deep learning system while considering the existence of unreliable participants, and presented a new scheme called SecProbe.
%Our scheme allows participants to collaboratively learn a collective deep learning model and gain benefits from the full data without needing to directly share and centrally collect their own data.
SecProbe utilizes exponential mechanism and functional mechanism to protect both the privacy of the participants' data and the quality of their data, the two major privacy concerns in such a system. The experimental results demonstrated that SecProbe is robust to unreliable participants, and can achieve high-accuracy results which are close to the model trained in a traditional centralized manner, while providing rigorous privacy guarantee.

\bibliographystyle{IEEEtran}
\bibliography{DP}
%\ifCLASSOPTIONcaptionsoff
%  \newpage
%\fi

\begin{IEEEbiography}[{\includegraphics[width=1in,clip,keepaspectratio]{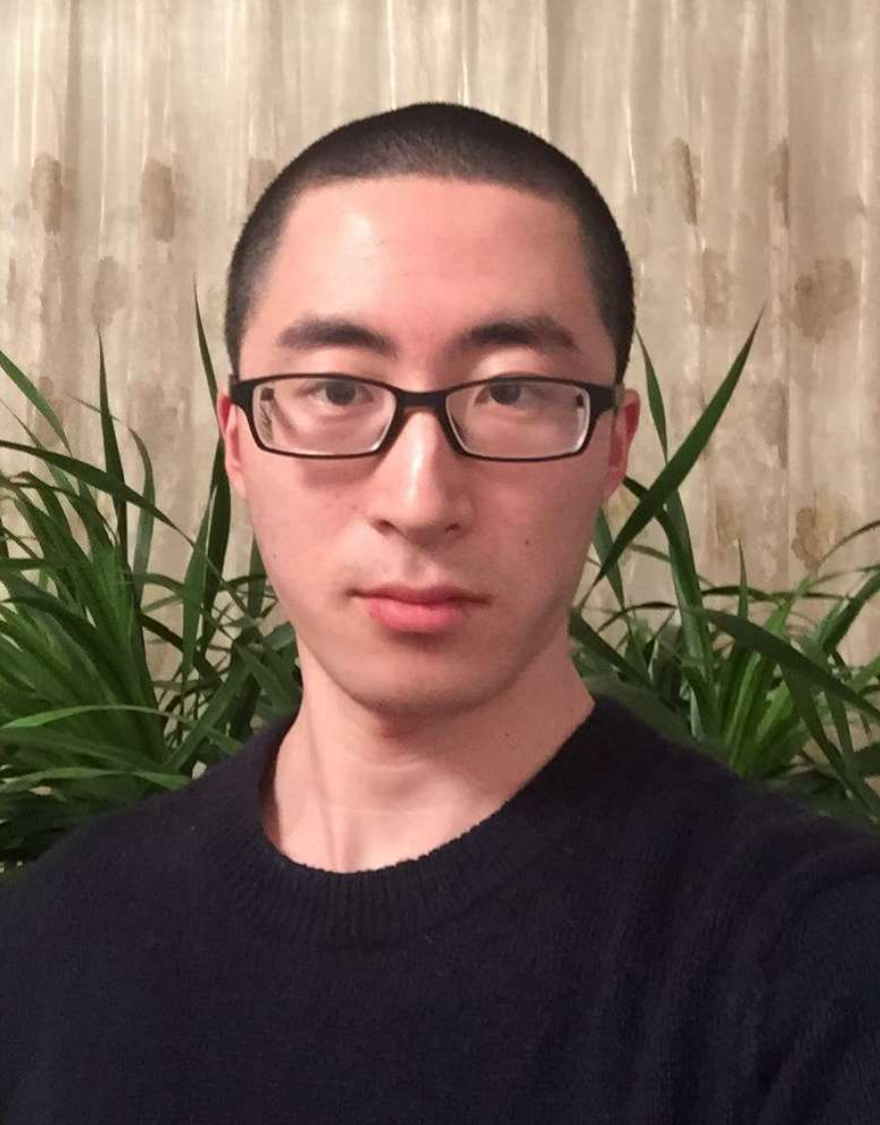}}]{Lingcheng Zhao}
received his B.S. degree from Central South University, China in 2016, in the College of Information Science and Engineering. He is currently pursuing the Ph.D. degree at the School of Cyber Science and Engineering, Wuhan University, China. His research interests include differential privacy and applied cryptography.
\end{IEEEbiography}
%\vspace{1mm}

\begin{IEEEbiography}[{\includegraphics[width=1in,clip,keepaspectratio]{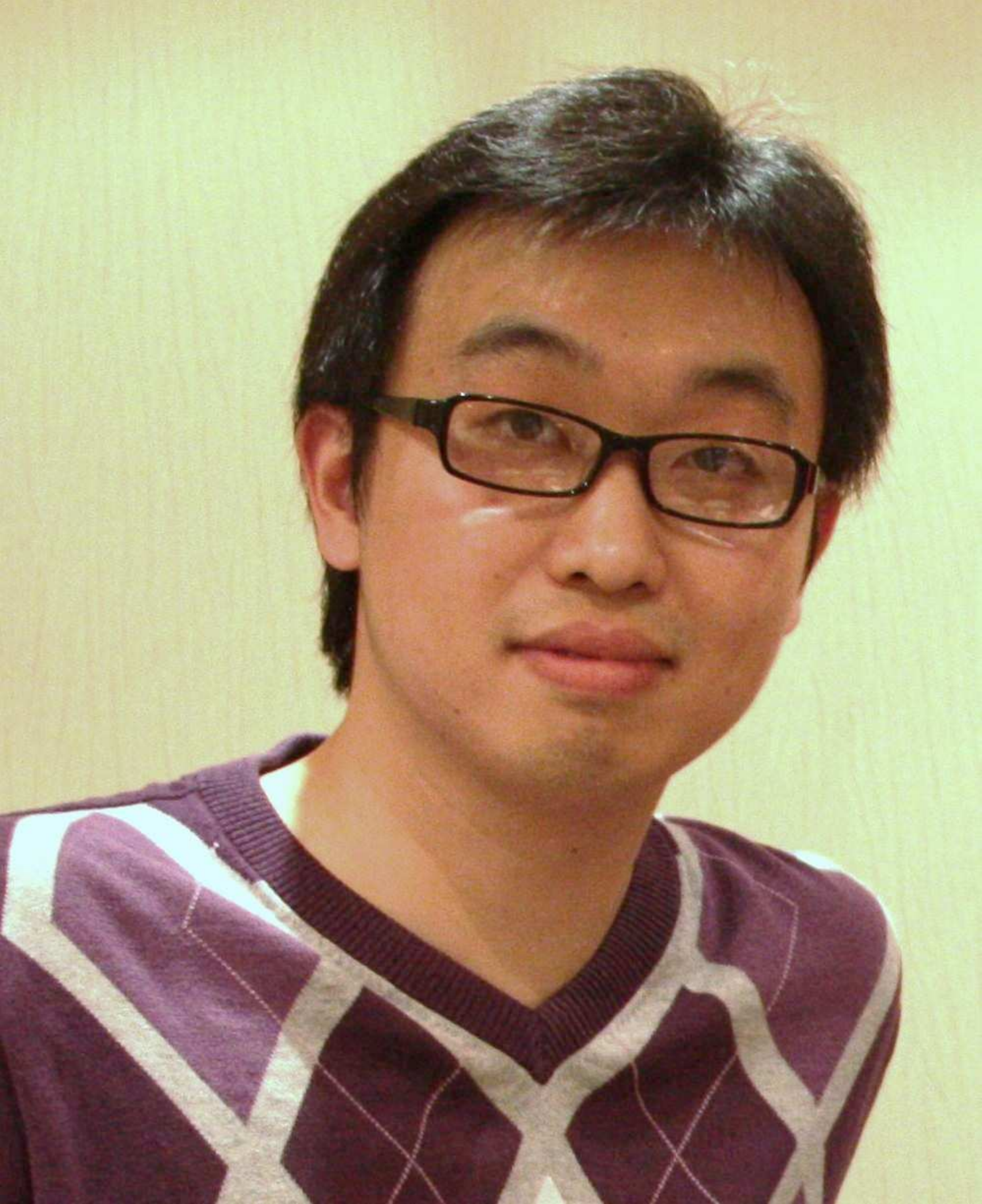}}]{Qian Wang}
is a Professor with the School of Cyber Science and Engineering, Wuhan University. He received the Ph.D. degree from Illinois Institute of Technology, USA. His research interests include AI security, data storage, search and computation outsourcing security and privacy, wireless system security, big data security and privacy, and applied cryptography etc. Qian received National Science Fund for Excellent Young Scholars of China in 2018. He is also an expert under National ``1000 Young Talents Program'' of China. He is a recipient of the 2016 IEEE Asia-Pacific Outstanding Young Researcher Award. He is also a co-recipient of several Best Paper and Best Student Paper Awards from IEEE ICDCS'17, IEEE TrustCom'16, WAIM'14, and IEEE ICNP'11 etc. He serves as Associate Editors for IEEE Transactions on Dependable and Secure Computing (TDSC) and IEEE Transactions on Information Forensics and Security (TIFS). He is a Senior Member of the IEEE and a Member of the ACM.
\end{IEEEbiography}

%\vspace{1mm}

\begin{IEEEbiography}[{\includegraphics[width = 1in,height = 1.25in,clip,keepaspectratio]{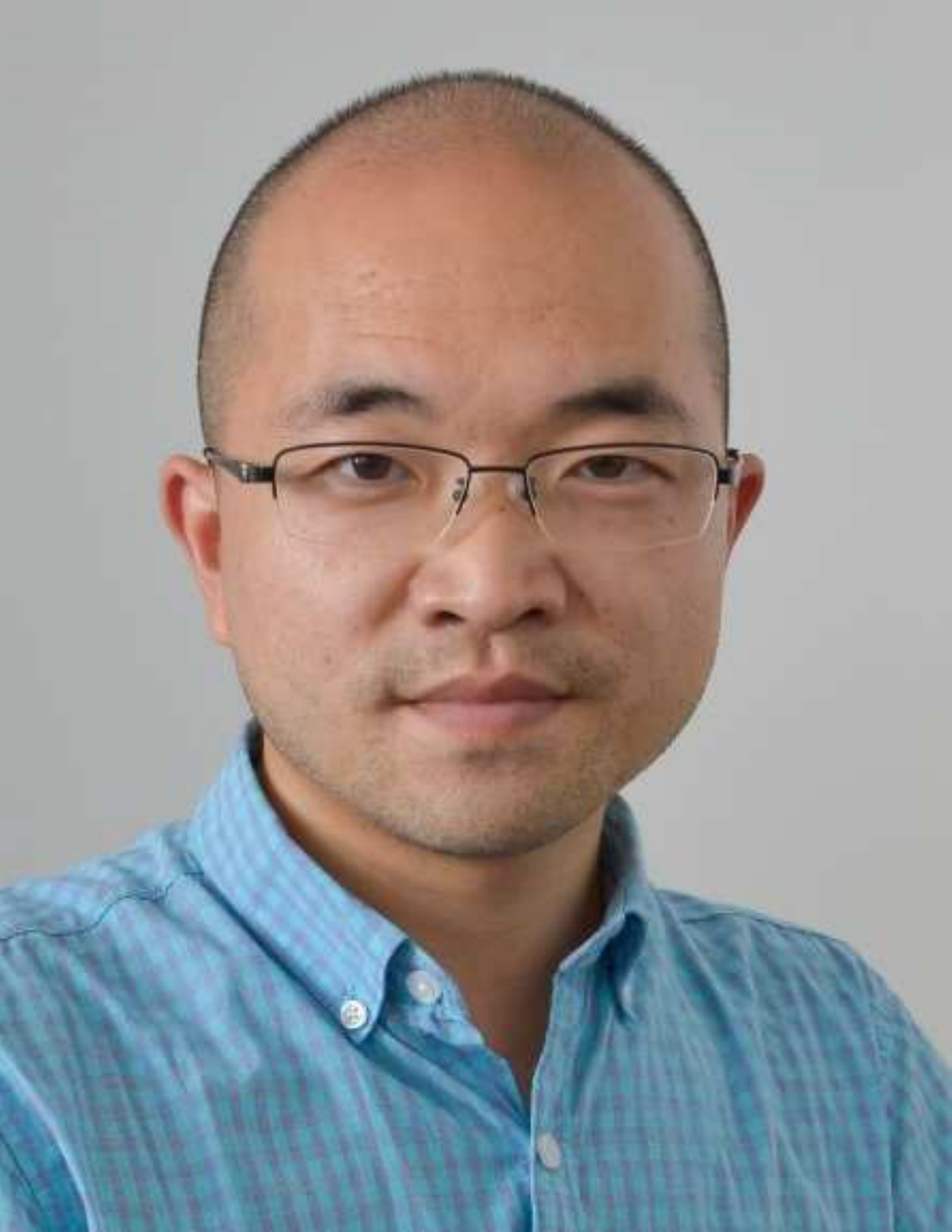}}]{Qin Zou}
received the B.E. degree in information engineering and the Ph.D. degree in computer vision from Wuhan University, China, in 2004 and 2012, respectively. From  2010  to  2011, he was a visiting PhD student at the Computer Vision Lab, University of South Carolina, USA. Currently, he is an Associate Professor at the School of Computer Science, Wuhan University. He is a co-recipient of the National Technology Invention Award of China 2015. His research activities involve computer vision, pattern recognition, and machine learning. He is a Senior Member of the IEEE and a Member of the ACM.
\end{IEEEbiography}

\begin{IEEEbiography}[{\includegraphics[width=1in,clip,keepaspectratio]{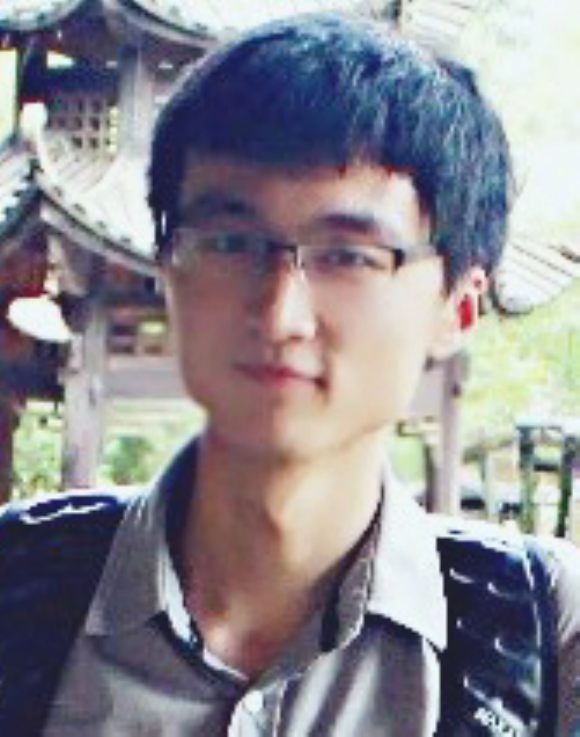}}]{Yan Zhang}
received the B.S. degree from China University of Geoscience in 2014 and the M.S. degree from Wuhan University in 2017. He is currently a software engineer at the Huawei Technologies Co., Ltd. His current research interests include differential privacy in data publishing, data mining and machine learning.
\end{IEEEbiography}
%\vspace{1mm}

\begin{IEEEbiography}[{\includegraphics[width=1in,clip,keepaspectratio]{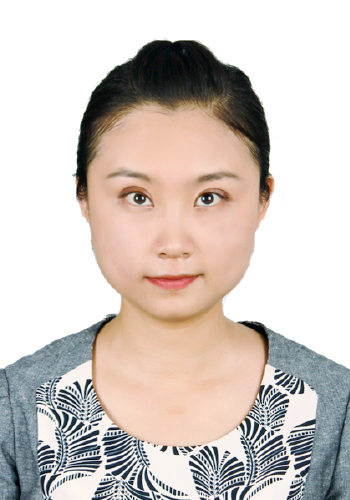}}]{Yanjiao Chen}
received her B.E. degree in Electronic Engineering from Tsinghua University in 2010 and the Ph.D. degree in Computer Science and Engineering from Hong Kong University of Science and Technology in 2015. She is currently a Professor in Wuhan University, China. Her research interests include spectrum management for Femtocell networks, network economics, network security, and Quality of Experience (QoE) of multimedia delivery/distribution. She is a Member of the IEEE and a Member of the ACM.
\end{IEEEbiography}
\vspace{1mm}

\ifCLASSOPTIONcaptionsoff
  \newpage
\fi
\end{document}